\documentclass{article} 
\usepackage[utf8]{inputenc}
\usepackage{graphicx}
\usepackage[usenames,dvipsnames,table,xcdraw]{xcolor}
\setkeys{Gin}{width=\columnwidth}
\usepackage{hyperref}
\usepackage[mode=buildnew]{standalone}
\usepackage{tikz}
\usepackage{circuitikz}
\usepackage{subcaption}
\usepackage{float}
\usepackage[ruled,vlined, linesnumbered]{algorithm2e}
\usepackage[backend=biber, url = false]{biblatex}
\title{The spring bounces back: Introducing the Strain Elevation Tension Spring embedding algorithm for network representation}

\author{Jonathan Bourne}
\date{July 2020}

\begin{document}

\maketitle

\begin{abstract}

This paper introduces the Strain Elevation Tension Spring embedding (SETSe) algorithm, a graph embedding method that uses a physics model to create node and edge embeddings in undirected attribute networks. Using a low-dimensional representation, SETSe is able to differentiate between graphs that are designed to appear identical using standard network metrics such as number of nodes, number of edges and assortativity. The embeddings generated position the nodes such that sub-classes, hidden during the embedding process, are linearly separable, due to the way they connect to the rest of the network. SETSe outperforms five other common graph embedding methods on both graph differentiation and sub-class identification. The technique is applied to social network data, showing its advantages over assortativity as well as SETSe's ability to quantify network structure and predict node type. The algorithm has a convergence complexity of around $\mathcal{O}(n^2)$, and the iteration speed is linear ($\mathcal{O}(n)$), as is memory complexity. Overall, SETSe is a fast, flexible framework for a variety of network and graph tasks, providing  analytical insight and simple visualisation for complex systems. 

\end{abstract}

\section{Introduction}

With the rise of social media and e-commerce, graph and complex networks have become a common concept in society. Their ubiquity and the already digitised nature of social networks has led to a great deal of research. Although there is a range of algorithms that can perform supervised learning directly on graphs \cite{ cao_deep_2016, kipf_semi-supervised_2016, seo_structured_2018, scarselli_graph_2009} (see \textcite{wu_comprehensive_2020} for a recent survey on the subject), a more common approach is to represent the nodes of the network in a latent vector space that traditional supervised learning techniques can then use.  These graph embeddings create a vector representation of the graphs, preserving valuable network properties such as distance on the graph, community structure and node class \cite{grover_node2vec_2016, ou_asymmetric_2016, perozzi_deepwalk_2014, roweis_nonlinear_2000}. These algorithms tend to find embeddings by minimising the distance between similar nodes, within the structure of the network. The literature review by \textcite{goyal_graph_2018} provides a survey of the most significant of these algorithms. With recent improvements in neural networks, there has been substantial growth in research on graph embedding algorithms, with over 50 new neural-network-based embedding algorithms published between 2015 and 2020 \cite{fey_fast_2019}.

Physics models can also be used to embed graphs in vector space; however, these are typically used only for drawing graphs. While other graph drawing techniques exist \cite{frick_fast_1995, koren_drawing_2005, krzywinski_hive_2012}, force-directed physics models are some of the most popular \cite{  eades_heuristic_1984, fruchterman_graph_1991, kamada_algorithm_1989}. These algorithms originated in the 1960s \cite{tutte_how_1963}, and use simple physics to find an arrangement of nodes that provides an aesthetically pleasing plot of a graph or network. These algorithms are sometimes called `spring embedders' \cite{kobourov_force-directed_2013} due to using springs or spring-like methods to place nodes and typically attempt to optimise an ideal distance between nodes, minimising the energy of the system. Although popular at the end of the 20th century, spring embedders became less common in research as machine learning became more popular.

The importance of drawing graphs is discussed in several papers \cite{chen_same_2018, matejka_same_2017, peel_multiscale_2018, revell_graphs_2018}. These researchers demonstrate that graphs that are structurally very different can appear identical until visualised. A popular statistical equivalent is Anscombe's quartet \cite{anscombe_graphs_1973}, a series of four figures showing very different data. However, in terms of the mean, variance, correlation, linear regression and ${R}^2$, all four figures in the quartet appear identical. It highlights the importance of visualising data and the weaknesses of some commonly used statistical tools.

The Strain Elevation Tension Spring embedding (SETSe) algorithm takes its name from the embeddings it produces. SETSe takes the node attributes of a graph, representing them as a force. The edges are represented by springs whose stiffness is dependent on the edge weight. The algorithm finds the position of each node on a manifold such that the internal forces created by the nodes are balanced by the resistive forces of the springs and the network is in equilibrium. The SETSe algorithm acts as a hybrid between the advanced techniques of the machine learning graph embedders used for analysis and the intuitive simplicity of the spring embedders used for graph drawing.

This paper demonstrates that in a world of sophisticated machine learning, there is still a role for simple, intuitive embedding methods. It shows that SETSe can find meaningful embeddings for the Peel's quintet \cite{peel_multiscale_2018} series of graphs, as well as the relations in Facebook data \cite{traud_social_2012}. It finds these embeddings efficiently in linear iteration time and space complexity. An R package has been created which provides all the functionality necessary to run SETSe analysis/embeddings (available from \url{https://github.com/JonnoB/rSETSe}).

The paper asks can SETSe distinguish between graphs that are identical using traditional network metrics? It also asks can SETSe be used to classify individual nodes? To gauge how well SETSe performs these tasks, it is compared against several popular graph embedding algorithms: node2vec \cite{grover_node2vec_2016}, SDNE \cite{wang_structural_2016}, LLE \cite{roweis_nonlinear_2000}, Laplacian Eigenmaps \cite{belkin_laplacian_2003} and HOPE \cite{ou_asymmetric_2016}.

\section{Method}
This  section begins by providing a simple example of the SETSe algorithm. It then describes and defines the physics model that underpins the embedding method. The algorithmic implementation and practical issues related to convergence are also briefly described. The datasets used in this paper are then introduced. Finally, the analyses performed are described.

\subsection{Introduction to SETSe: a simple example}
\label{sect:conceptual_example}
The SETSe algorithm takes a network $\mathcal{G}$ containing the set of V nodes and the set E edges. It converts the $n$ attributes or variables of each node into orthogonal forces where each attribute occupies a single dimension. 
In each dimension, the total sum (across the network) of the forces in that dimension is 0 (i.e. the network is balanced in all dimensions). Each edge is converted to a spring whose stiffness $k$ is taken from an attribute of the edge, typically the edge weight. If no edge attribute is to be used, all the edges in the network take $k$ as an arbitrary constant. The algorithm then positions each node on an $n+1$-dimensional manifold such that no node experiences a net force. Like other spring embedders \cite{fruchterman_graph_1991, kamada_algorithm_1989,quigley_fade:_2001}, SETSe is subject to the n-body problem \cite{aarseth_n-body_2003-1, springel_simulations_2005} and must be solved iteratively.

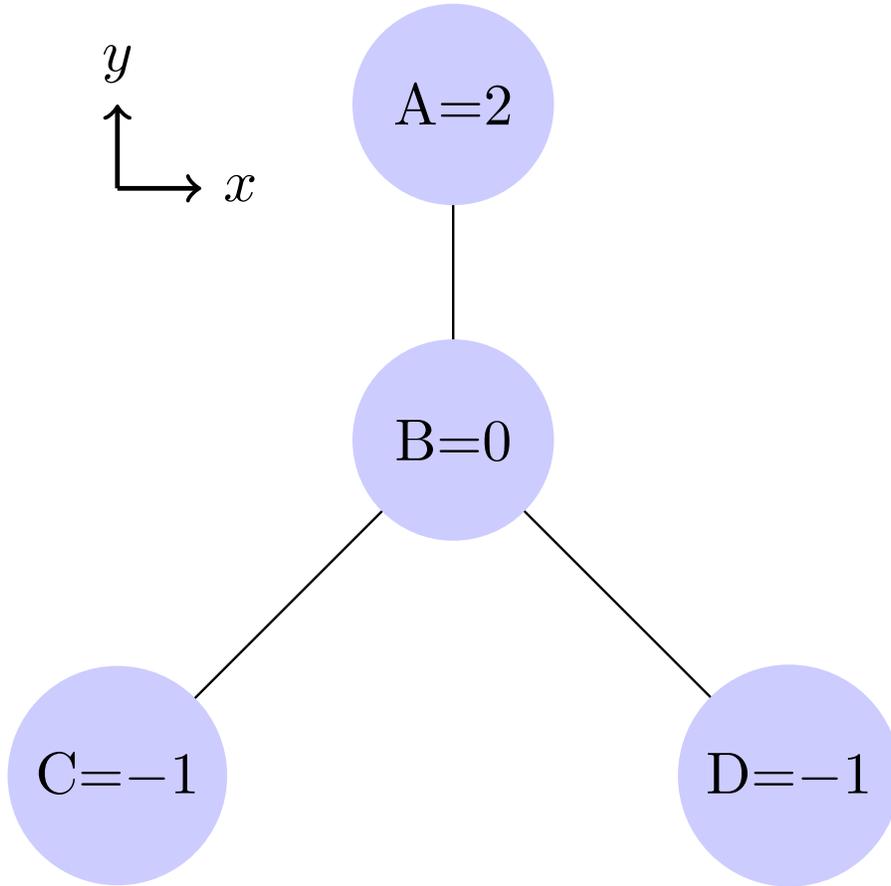
\begin{figure}[h!]
  \begin{center}
    \tikzset{blue_fill/.style={circle,fill=blue!20, minimum size=1.2cm, auto = left}}

\begin{tikzpicture}
    
    \draw[thick, ->] (-2,1.5) -- (-1.5,1.5) node[anchor = west] {$x$};
    \draw[thick, ->] (-2,1.5) -- (-2,2) node[anchor = south] {$y$};

        \draw
  (0,0)  -- node[right=1pt] {} (2,-2) coordinate (d) node[blue_fill] {D=$-1$};
    
  \draw
  (0,2) coordinate (a) node[blue_fill] {A=$2$}
  -- node[left=1pt] {} (0,0) coordinate (b) node[blue_fill] {B=0}
  -- node[left=1pt] {} (-2,-2) coordinate (c) node[blue_fill] {C=$-1$};
  
  \end{tikzpicture}
    \caption{A network of four nodes and three edges. The node attributes are considered forces that act in the $z$-direction. The forces acting on the network balance, but the forces acting on the nodes only balance when the nodes are in the appropriate position in the $z$-axis; the network is shown in the $x$--$y$ planes.}
    \label{tikz:nodes_4_edges_3}
  \end{center}
\end{figure}

The functioning of SETSe is best described using example. Consider the simple network shown in figure \ref{tikz:nodes_4_edges_3}. The network is planar, and so can be drawn in two dimensions ($x$ and $y$), with no edges crossing. The nodes in the network have a single attribute/variable that acts perpendicular to the plane; as such, the nodes in the network act as beads whose movement is restricted to the $z$-axis and are fixed in $x$ and $y$. The nodes in the network have an identical mass $m$, and the edges between the nodes have a common distance $d$, which is the length of the spring at rest. Node A exerts a force of 1, node B exerts no force on the network, while nodes C and D exert a force of $-1$ each, resulting in a net force of $0$.  The edges of the network are springs that, when stretched, act according to Hooke's law $F= \Delta Hk$, where $\Delta H$ is the extension of the edge and $k$ the spring stiffness such that $0 < k \leq \infty$.

Although the vertical forces (defined by the node attribute) across the network sum to 0, the individual nodes are not in equilibrium and so begin to move in the direction of their respective forces. The vertical distance between node pairs resulting from this movement extends the springs, creating a resistive force. The network will find a three-dimensional equilibrium when the net force acting on each node is 0. This occurs when the elevation of the nodes in the system is such that for each node, the sum of the vertical tension in all edges connected to that node is equal and opposite to the force produced by the node itself. As an example, if $k=1000$ and $d=1$, the equilibrium positions of the nodes are  0.1450, 0.0185, $-0.0818$ and $-0.0818$ for the nodes A to D, respectively. The interested reader can confirm that the net forces on each node are $0$, using Pythagoras' theorem and Hooke's law.

A crucial point in this example is that the nodes are beads. As such, the $xy$ positions are fixed and only the vertical component of the spring tension has an impact on the nodes; all horizontal forces can be disregarded. Ignoring the horizontal force means that the $xy$ position of the nodes can be ignored, reducing the initial layout of the network to a zero-dimensional point in space, out of which a one-dimensional node elevation appears. The horizontal distance between nodes is reduced to a crucial but abstract mapping value.

\begin{figure}
    \centering
    \includegraphics{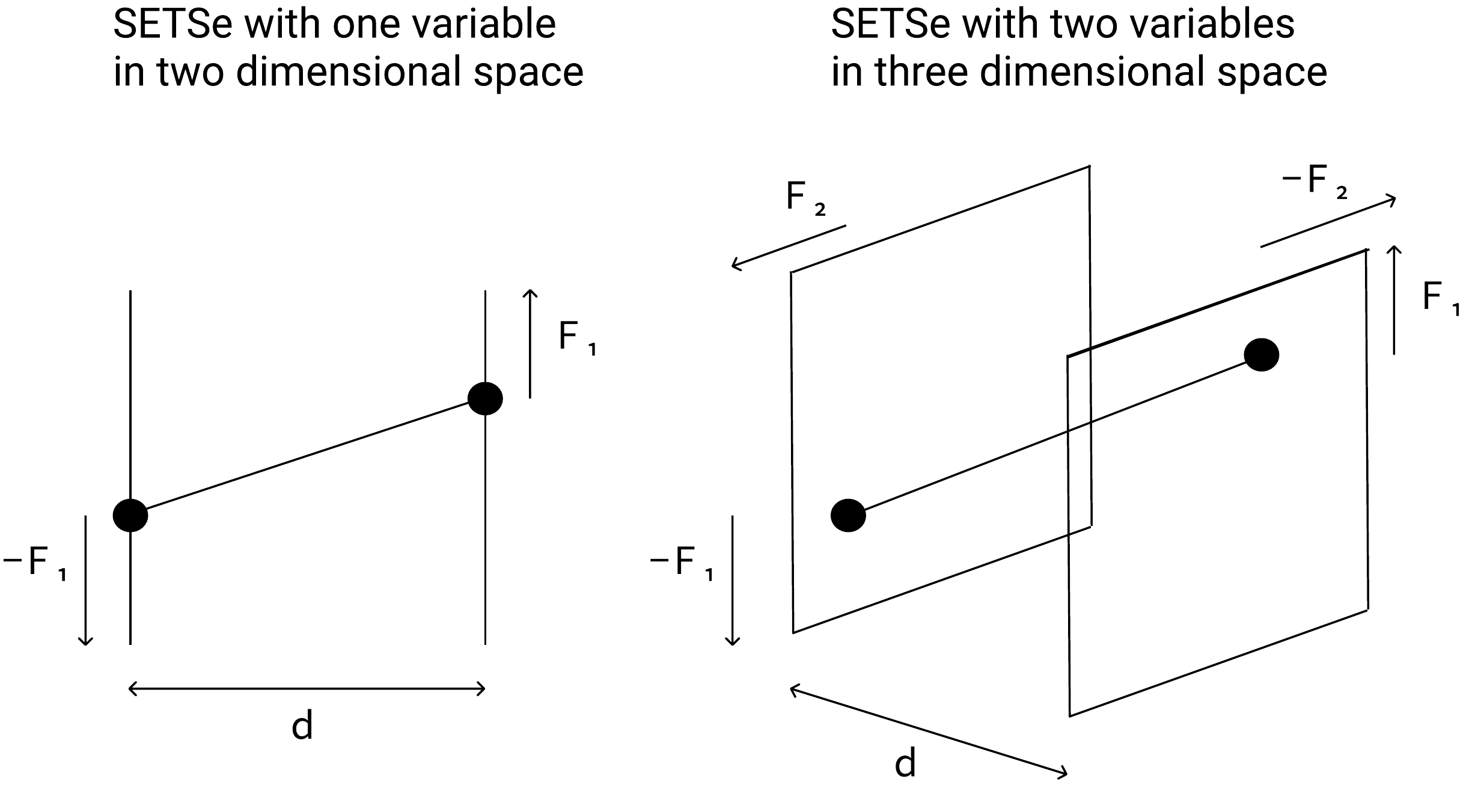}
    \caption{Two nodes in a network in two-dimensional and three-dimensional SETSe systems. The minimum distance $d$ between the nodes is maintained by the graph space dimension.}
    \label{fig:2d-3d_setse}
\end{figure}

SETSe space is a non-Euclidean metric space of $n+1$ dimensions where $n$ is the number of attributes the network has. The $n+1$th dimension is the graph space, representing the graph adjacency matrix. The distance between connected nodes in the graph space is $d_{i,j}$. Dimensions 1 to $n$ are Euclidean, while the graph dimension is not. This concept is visualised in figure \ref{fig:2d-3d_setse}, which shows nodes embedded in pairwise two-dimensional space and pairwise three-dimensional space. The graph space acts as the minimum distance between the nodes. Figure \ref{fig:2d-3d_setse} shows that the nodes occupy parallel Euclidean hyper-planes separated by the graph space. As SETSe space is locally Euclidean, it is an $n+1$-dimensional manifold. As an example of a network that is pairwise Euclidean but not Euclidean overall, consider a maximally connected network of four nodes. There is no arrangement of the nodes on a plane where the distance between all nodes is equal, although pairwise all nodes can have the same distance.

\subsection{Creating the physics model}
The calculation of the solution of a single variable graph is described below. The extension for higher dimensions is described in the subsequent paragraph.
The net force acting on node $i$ can be written as $F_{\textrm{net}, i} = F_{i} - F_{\mathrm{vten,i}}$, where $F_{i}$ is the force produced by the node and $F_{\mathrm{vten,i}}$ is the vertical component of the net tension acting on the node from the springs. The net force on node $i$ is shown again in equation \ref{eq:equilibrium} where $F_\mathrm{ten, i,j}$ is the total tension in edge $i,j$. The angle $\theta_{i,j}$ is the angle of the force between nodes $i$ and $j$. The tension in an edge is given by Hooke's law as $F_\mathrm{ten, i,j}=k_{i,j}(H_{i,j}-d_{i,j})$ where $H_{i,j}$ is the length of the extended spring and $d_{i,j}$ is the graph distance between the nodes. The length of the extended spring length $H_{i,j}$ can be found, as it is the hypotenuse of the distance triangle between nodes $i$ and $j$ such that $H_{i,j} = \sqrt{\Delta z_{i,j}^2 + d_{i,j}^2}$, where $\Delta z_{i,j}$ is the elevation difference between nodes $i$ and $j$, $\Delta z_{i,j} = z_j- z_i$. As $\mathrm{cos} \, \theta = \frac{\Delta z}{H}$, the equation for net tension can be rearranged into an alternative expression of edge tension, which is shown in equation \ref{eq:edge_ten3}. The strain component of SETSe is simple mechanical strain and is shown in equation \ref{eq:strain}. Strain and tension are perfectly correlated in the special case that $k$ is constant for all edges in the network.

\begin{equation}
    F_{\mathrm{net},i} =F_i - \sum_j^n F_{\mathrm{ten},i,j} \,\mathrm{cos} \, \theta_{i,j}
    \label{eq:equilibrium}
\end{equation}

\begin{equation}
    F_{\mathrm{net},i} =F_i -  \sum_j^n k_{i,j}(H_{i,j}-d_{i,j}) \frac{\Delta z_{i,j}}{H_{i,j}}
    \label{eq:edge_ten2}
\end{equation}

\begin{equation}
    F_{\mathrm{net},i} =F_i -  \sum_j^n k_{i,j} \Delta z_{i,j} (1-\frac{d_{i,j}}{H_{i,j}})
    \label{eq:edge_ten3}
\end{equation}

\begin{equation}
    \varepsilon_{i,j} = \frac{H_{i,j}-d_{i,j}}{d_{i,j}}
    \label{eq:strain}
\end{equation}

The extension of SETSe from a graph with a single attribute to a graph with $n$ attributes is straightforward. 
The hypotenuse vector is $\textbf{H}_{i,j} = \textbf{z}_{i}-\textbf{z}_{j} + \textbf{d}$, which is a vector of $n+1$ elements, where the first $n$ elements are the differences in position between nodes $i$ and $j$ in the $n$ dimensions, and the $n+1$th dimension is the graph distance $d$. The scalar length of $\textbf{H}_{i,j}$ is the Euclidean distance between the nodes in $n+1$-dimensional space i.e. $H_{i,j}= \sqrt(\sum^n_1 (z_{i,q}-z_{j,q})^2 + d_{i,j}^2)$. To find the angle between the hypotenuse and the distance between the two nodes in dimension $q$, the cosine similarity is used, as shown in equation \ref{eq:cosine}. As all entries of $\Delta \textbf{z}_{q,i,j}$ are 0, apart from the $q$th entry, the cosine similarity simplifies to equation \ref{eq:cosine_simple}, which is the `vertical' distance between the nodes in dimension $q$ over the scale length of $H_{i,j}$. It is then easy to see that equation \ref{eq:tension_comp_multi} is the multidimensional equivalent of equation \ref{eq:edge_ten3}.

\begin{equation}
     \textrm{cos} \; \theta_{q,i,j} = \frac{\textbf{z}_{q,i,j} \cdot \textbf{H}_{i,j}}{\left \| \textbf{z}_{q,i,j} \right \| \left \| \textbf{H}_{i,j} \right \|}
     \label{eq:cosine}
\end{equation}

\begin{equation}
    \textrm{cos} \; \theta_{q,i,j} = \frac{\Delta z_{q,i,j} }{H_{i,j}}
         \label{eq:cosine_simple}
\end{equation}

\begin{equation}
       F_{\textrm{net},q, i}= F_{q,i} - \sum_j^n k_{i,j}\Delta z_{q,i,j}(1-\frac{d_{i,j}}{H_{i,j}})
    \label{eq:tension_comp_multi}
\end{equation}

The distance $d$ between the nodes is a key parameter when it comes to finding the final elevation embedding. If the distance is not a constant, it must be meaningful for the type of network being analysed. As an example, the distance in metres between two connected points on an electrical circuit is unlikely to be meaningful; however, a variable distance may be appropriate if traffic were being analysed.
Understanding meaningful distance metrics is not explored in this paper, and distance between nodes is considered a constant across all edges. 

SETSe can be used on continuous and categorical variables. In both cases, the forces must be balanced; this is when the net value of the sum of the forces across all nodes equals 0. For continuous variables, the raw attribute force of node $i$ ($F_i$) is normalised to create balanced force $F_{i,\mathrm{bal}}$ by subtracting the mean from all values, as shown in equation \ref{eq:cont_var}, where  $\left | \mathcal{V} \right |$ is the number of nodes in the network.

\begin{equation}
    F_{i,\mathrm{bal}} = F_i - \frac{1}{\left | \mathcal{V} \right |} \sum_{1}^{\left | \mathcal{V} \right |} F_i
    \label{eq:cont_var}
\end{equation}

For categorical node attributes, each level is treated as a binary attribute, making as many new dimensions as there are levels; this is similar to how variables in linear and logistic regression are treated. The total force level dimension $\gamma$ is the fraction that level makes up of the total number of nodes, as shown in equation \ref{eq:for_level}. The force produced by the nodes in each level dimension can then be treated as continuous variables, as described previously in equation \ref{eq:cont_var}.

 \begin{equation}
     F_{\gamma} =  \frac{\left | \mathcal{V}_\gamma  \right |}{\left | \mathcal{V}\right |}
     \label{eq:for_level}
 \end{equation}

With the force and distance relationship between pairs of nodes defined, it is now possible to look at the method used to find the equilibrium state of the network and its corresponding strain, elevation and tension embeddings. The difficulty in solving such a problem is that the relationship between the final elevation of a node and the force it experiences is non-linear. In addition, each node is affected by all other nodes and spring stiffness $k$ in the network. This interaction creates a situation that is similar to the n-body problem of astrophysics. In this case, although the nodes act as bodies, instead of exerting a force on all other nodes, as celestial bodies do, they only exert a force on those nodes with which they have a direct connection. 

The equilibrium solution can be found by treating the problem as a dynamic system and iterating through discrete time steps until the system reaches the equilibrium point. By representing the network as a dynamic system, the acceleration and velocity of each node must be calculated. Using Newton's second law of dynamics, $F=ma$, where $F$ is the force acting on the node and $a$ is the acceleration, the nodes need to be assigned an arbitrary constant mass $m$ (note mass does not affect the final embeddings). The net force acting on each node at time step $t$ is then equation \ref{eq:NetForce}, where $F$ is the force generated by the node according to the node attribute. The system is assumed to be a viscous laminar fluid, and so the damping is simply the product of the velocity $v$ and the coefficient of drag $c$. Friction is used to cause the system to slowly lose energy and converge. While it does not affect the value at convergence, it needs to be correctly parametrised or the system will not converge. This is discussed in the Appendix.

\begin{equation}
     F_\mathrm{net,i}= F_{\mathrm{vten}_i} + F_i- c v_i
     \label{eq:NetForce}
\end{equation}

Knowing the net force acting on the node allows calculation of the equations of motion at each time step. Velocity can be calculated as $v_{t} = v_{t-1} + \frac{F_{\mathrm{net},t}}{m}\Delta t$, where $v$ is the velocity at time $t$. Distance is the elevation embedding and is calculated by $z = v_{t-1} \Delta t + \frac{1}{2}a_t \Delta t^2 + z_{t-1}$.

The SETSe algorithm is shown in algorithm \ref{algo:find_Stable}. The equations described in equations \ref{eq:equilibrium} to \ref{eq:NetForce} are either converted to vectors or matrices, allowing all nodes and edges in the network to be updated simultaneously. The algorithm takes a graph $\mathcal{G}$, which has been processed so that each edge has distance $d_{i,j}$ and spring constant $k_{i,j}$. The dynamics of the network are all initialised at $0$; only the forces exerted by the nodes are non-zero values. In algorithm \ref{algo:find_Stable}, vectors are lower case letters in bold, while matrices are in bold and capitals. 
The time in the system is represented by $t$ and the time step per iteration is $\Delta t$. The elevation of each node in the system is represented by the matrix $\textbf{Z}$. The elevation difference across each edge is $\Delta \textbf{Z}$, and is obtained by subtracting the transpose of the elevation matrix from the original elevation matrix. The hypotenuse, or total length, of the edge is represented by the matrix $\textbf{H}$ and is found using the elevation difference $\Delta \textbf{Z}$ as well as the horizontal difference $\textbf{d}$. The vertical component of the tension in each edge is represented by $\textbf{F}_{vten}$ and is the element-wise product of the edge spring stiffness matrix $\textbf{K}$ with the extension of the edge; this matrix is then multiplied element-wise again using the element-wise product by the tangent of the angle of the edge. Line 8 shows that the vertical component of the force $\textbf{f}_{vten}$ is updated by summing the rows of each line in the $\textbf{F}_{vten}$ matrix using a column vector of 1s, that is, $\left | \mathcal{V} \right |$ long. The elevation of each node is updated on line 9 of the algorithm. The vector $\textbf{z}$ is then reshaped using a function into matrix form.  Line 11 updates the velocity of each node in the network. Line 12 updates the static force on each node $\textbf{f}_{static}$. Static force is the force exerted by the node minus the sum of the tensions exerted by all the connected edges.  The next update is the system friction or drag $\textbf{f}_d$. The system force $\textbf{f}_{net}$ is then updated. Finally, the acceleration $\textbf{a}$ to be used in the next iteration is calculated.

One of the advantages that SETSe has over traditional force expansion algorithms \cite{kamada_algorithm_1989, fruchterman_graph_1991} is that the distance from the optimal solutions is known. In the other algorithms, the loss function is to reduce the total energy of the system to some unknown minimum. However, SETSe has a loss function more similar to the error metrics used in statistics or machine learning. The ideal static force of the system is $0$ and the initial static force is $\sum \left \| F_i \right \| $, which is thus bounded in a finite space. This is an important consideration when it comes to efficient convergence and auto-convergence and is discussed further in the Appendix.
Although the stop condition of the algorithm is that the static force in the network is 0, in practice, the system is said to have converged if $\textbf{f}_{static} \approx 0$. In this paper, the tolerance for convergence will be $f_{static} \leq \frac{ \sum \left \| F_i \right \| }{10^3}$, that is when the static force is reduced to 1/1000th of the absolute sum of forces exerted by the nodes.

\begin{algorithm}
\SetAlgoLined
\KwResult{The strain, elevation and tension embeddings of the original graph}
 $t = 0$ \;
$\textbf{f}_{static} = \textbf{f}$\;
 \While{$ \sum \left \|\textbf{f}_{static} \right \| \neq 0$ }{
 $t = t +\Delta t$\;
 $\Delta \textbf{Z} = \textbf{Z} - \textbf{Z}^T$\;
$\textbf{H} = \sqrt{\Delta \textbf{Z}^2 +\textbf{d}^2}$\;
$\textbf{F}_{vten} = \textbf{K}\circ(\textbf{H} -\textbf{d})\circ\frac{\Delta \textbf{Z}}{\textbf{H}}$\;
$\textbf{f}_{vten}  = \textbf{F}_{vten} \textbf{J}_{\left | \mathcal{V}  \right |}$\;
 $\textbf{z} = \textbf{v}\Delta t + \frac{1}{2}\textbf{a}\Delta t^2 + \textbf{z}$\;
$\textbf{Z}= \textbf{f(z)}$\;
 $\textbf{v} = \textbf{v}+ \textbf{a}\Delta t$\;
$\textbf{f}_{static} = \textbf{f}-\textbf{f}_{vten}$\;
 $\textbf{f}_d = c_d \textbf{v}$\;
$\textbf{f}_{net} = \textbf{f}_{static}-\textbf{f}_d$\;
 $\textbf{a} = \frac{\textbf{f}_{net}}{m}$\;
 }

 \caption{The SETSe algorithm}
 \label{algo:find_Stable}
\end{algorithm}

\subsubsection{Practical convergence issues}
The implementation of the algorithm in the R package \texttt{rSETSe} has two modes: sparse and semi-sparse. Semi-sparse  mode is used on smaller graphs, and sparse mode is for larger graphs (starting at 5000--10,000 edges), the complexity of sparse mode is  linear to the number of edges $\mathcal{O}(\left | \textrm{E} \right |)$. The mode has no impact on the final embeddings. Time and space complexity are discussed further in section \ref{sect_complexity}.

Although there are several parameters of the physical model that must be initialised before running the algorithm, only two have any real bearing on the final embeddings. Drag, time step and mass affect the rate of convergence, but not the final outcome (when $F_\textrm{static} = 0$). The distance between the nodes does affect the outcome as final elevation will change. However, when $F_\textrm{static} = 0$, the angle between the nodes is unaffected by the length of the edges and so elevation can be normalised. Only the force variable and the spring stiffness have an impact on the final converged values. The force variable is not controlled by the user, leaving only $k$. As such, the value of $k$ must be constant for all networks under evaluation, or if $k$ is a function, then $k=\textrm{f}(x)$ must be consistently parametrised. 

All networks in this paper are embedded using bi-connected SETSe, a more advanced method than algorithm \ref{algo:find_Stable}. This method breaks the network into bi-connected sub-graphs then calls auto-SETSe, which is an algorithm that chooses the coefficient of drag using a binary search.  Both bi-connected SETSe and auto-SETSe are discussed in detail in the Appendix.

\subsection{Data}
Two datasets are used in this paper to illustrate how SETSe works and how it can be used to gain insight into network structure and behaviour. Both of the datasets have binary edges, meaning that $k$ is constant across all edges in all networks. This reduces the embeddings from three to two, as strain and tension have a perfect linear relationship.

\subsubsection{Peel's quintet}
\begin{figure}
    \centering
    \includegraphics{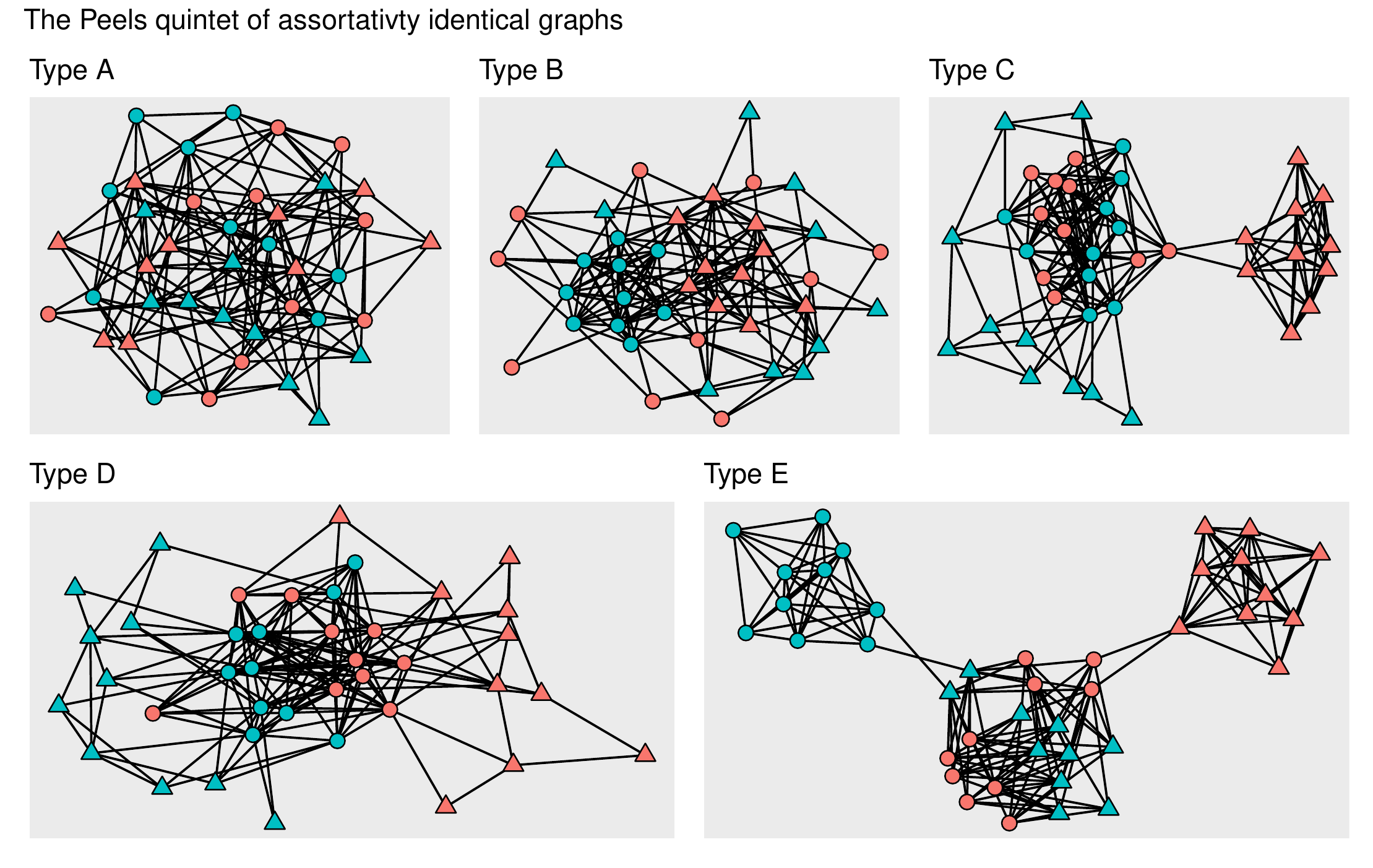}
    \caption{These networks were introduced by \textcite{peel_multiscale_2018}. They are all networks that have identical numbers of nodes edges, group size, within-group connections and between-class connections. However, the networks are clearly structurally distinct.}
    \label{fig:peels_quintet}
\end{figure}

Peel's quintet \cite{peel_multiscale_2018} is an example of the graph equivalent of Anscombe's quartet \cite{anscombe_graphs_1973}. It is a collection of five binary attribute graphs that have an identical number of nodes, edges connections between and within classes, and assortativity. The networks are very different when visualised (see figure \ref{fig:peels_quintet}). \textcite{peel_multiscale_2018} achieve this situation by dividing the binary classes into two sub-classes, which have different mixing patterns. They then develop an alternative metric and demonstrate that it can distinguish between the quintet and other network structures. Peel's quintet is essentially a hierarchical stochastic block model. Each network has two blocks containing two sub-classes. Each sub-class contains 10 nodes, with a total of 40 nodes per network. Each network has 120 edges with 40 edges connecting the classes together and 40 edges internally in each class. Because the number of edges connecting within and between the sub-class is distinct, the overall network structure is itself distinct, even though in terms of traditional network metrics they are identical. Peel's quintet will be used as an example of how SETSe is affected by graph topology and the network attributes, in this case, the two known communities and the hidden communities.

Figure \ref{fig:peels_quintet} shows Peel's quintet. The classes are shown as being either turquoise or red, while the two hidden classes are triangles or circles. While type A is simply a random network, the other networks show varying types of structure produced by the inter hidden group connection patterns. Table \ref{tab:peels} shows the block models the network is based on.

\setlength\tabcolsep{2pt}
\begin{table}[!htb]
    \caption{Block model for Peel's quintet}
    \begin{subtable}{.3\linewidth}
      \caption{Type A}
      \centering
\begin{tabular}{l|ll|ll}
\textbf{}   & \textbf{a1} & \textbf{a2} & \textbf{b1} & \textbf{b2} \\ \hline
\textbf{a1} & 10          & 20          & 20          & 20          \\
\textbf{a2} &   -         & 10          & 20          & 20          \\ \hline
\textbf{b1} & -           & -           & 10          & 20          \\
\textbf{b2} & -           & -           & -           & 10         
\end{tabular}
    \end{subtable}%
    \begin{subtable}{.3\linewidth}
      \centering
        \caption{Type B}
\begin{tabular}{l|ll|ll}
\textbf{}   & \textbf{a1} & \textbf{a2} & \textbf{b1} & \textbf{b2} \\ \hline
\textbf{a1} & 38          & 2           & 20          & 20          \\
\textbf{a2} & -           & 0           & 20          & 20          \\ \hline
\textbf{b1} & -           & -           & 0           & 2           \\
\textbf{b2} & -           & -           & -           & 38         
\end{tabular}
    \end{subtable}%
    \begin{subtable}{.3\linewidth}
      \centering
        \caption{Type C}
\begin{tabular}{l|ll|ll}
\textbf{}   & \textbf{a1} & \textbf{a2} & \textbf{b1} & \textbf{b2} \\ \hline
\textbf{a1} & 38          & 2           & 0           & 0           \\
\textbf{a2} & -           & 0           & 80          & 0           \\ \hline
\textbf{b1} & -           & -           & 10          & 20          \\
\textbf{b2} & -           & -           & -           & 10         
\end{tabular}
    \end{subtable}%
    \\
    \begin{subtable}{.5\linewidth}
      \centering
        \caption{Type D}
\begin{tabular}{l|ll|ll}
\textbf{}   & \textbf{a1} & \textbf{a2} & \textbf{b1} & \textbf{b2} \\ \hline
\textbf{a1} & 10          & 20          & 0           & 0           \\
\textbf{a2} & -           & 10          & 80          & 0           \\ \hline
\textbf{b1} & -           & -           & 10          & 20          \\
\textbf{b2} & -           & -           & -           & 10         
\end{tabular}
  \end{subtable}%
    \begin{subtable}{.5\linewidth}
      \centering
        \caption{Type E}
\begin{tabular}{l|ll|ll}
\textbf{}   & \textbf{a1} & \textbf{a2} & \textbf{b1} & \textbf{b2} \\ \hline
\textbf{a1} & 38          & 2           & 0           & 0           \\
\textbf{a2} & -           & 0           & 80          & 0           \\ \hline
\textbf{b1} & -           & -           & 0           & 2           \\
\textbf{b2} & -           & -           & -           & 38         
\end{tabular}
    \end{subtable} 
    \label{tab:peels}
\end{table}

\subsubsection{Facebook data}
The Facebook 100 dataset by \textcite{traud_social_2012} is a snapshot of the entire Facebook network on a single day in September 2005. At this time, Facebook was  open only to 100 US universities. There were very few links between universities then, so each one can be considered a stand-alone unit. Such an assumption would not be possible now. The data allow insight into the structure of relationships in attributed social networks. The networks are anonymised and the universities are referred to by a reference. Caltech36 is the smallest university network and has only 769 nodes and 16,656 edges, while Penn94 has the most nodes with 41,554 and Texas84 has the most edges with 1,590,655. The original study on this dataset \cite{traud_social_2012} found there were assortativity patterns within the variables that were generally common across all universities, such as tendency to be connected to students who will graduate at the same time or who lived in the same university accommodation.  The networks have seven attributes, all of which are categorical and have been anonymised. These attributes are: student type, gender, major, minor, dorm, year of graduation and high school.

\subsection{Experimental analysis}

The experiments are broken across the two datasets. The first set of experiments will focus on Peel's quintet, distinguishing network types, then distinguishing between node types. The second set of experiments will look at the Facebook data, first comparing assortativity with SETSe, which is similar to distinguishing between networks, then distinguishing between node types on the Facebook dataset. The two SETSe dimensions will be elevation and node tension. Node tension is the mean absolute tension in the edges connected to the nodes $ { v_\textrm{ten,i} = \frac{\sum F_{\textrm{ten},i,j}}{n} }$, where for this expression only, $n$ is the number of edges for node $v_i$. 

This paper uses four accuracy metrics: accuracy (eq \ref{eq:ACC}), balanced accuracy (eq \ref{eq:BAL_ACC}), f1 score (eq \ref{eq:f1}) and Cohen's kappa (eq \ref{eq:kappa}), where $\textrm{P}$, $\textrm{N}$, $\textrm{TP}$, $\textrm{TN}$, $\textrm{FP}$ and $\textrm{FN}$ are the number of positives, negatives, true positives, true negatives, false positives and false negatives, respectively. TPR is the false positive rate, $\textrm{TPR}= \frac{\textrm{TP}}{P}$, and TNR is the true negative rate  $\textrm{TNR}= \frac{\textrm{TN}}{N}$. $p_o$ is the observed probability of two events occurring, e.g. predict class one truth is class one. $p_e$ is the expected probability of two events occurring, given their overall prevalence. 

\begin{equation}
    \textrm{ACC} = \frac{\textrm{TP} + \textrm{TN}}{\textrm{P}+\textrm{N}}
     \label{eq:ACC}
\end{equation}

\begin{equation}
    \textrm{BAL\_ACC} = \frac{\textrm{TPR} + \textrm{TNR}}{2}
         \label{eq:BAL_ACC}
\end{equation}

\begin{equation}
    \textrm{f1} = \frac{2\textrm{TP} }{\textrm{TP} +\textrm{FP} +\textrm{FN} }
         \label{eq:f1}
\end{equation}

\begin{equation}
    \kappa = \frac{p_o - p_e}{1-p_e}
    \label{eq:kappa}
\end{equation}

\subsubsection{Distinguishing between networks}
The first analysis of the performance of SETSe will be to compare it to a selection of other node embedding methods using Peel's quintet \cite{peel_multiscale_2018}. The methods that it will be compared against are  node2vec \cite{grover_node2vec_2016}, SDNE \cite{wang_structural_2016}, LLE \cite{roweis_nonlinear_2000}, Laplacian Eigenmaps \cite{belkin_laplacian_2003} and HOPE \cite{ou_asymmetric_2016}. The methods cover three main areas: graph factorisation \cite{roweis_nonlinear_2000, belkin_laplacian_2003,ou_asymmetric_2016}, random walks \cite{grover_node2vec_2016} and deep learning \cite{wang_structural_2016}. A set of 100 networks from each of the five classes of Peel's quintet will be generated and embedded into two dimensions using each of the embedding methods. The embeddings are produced at node level and will be aggregated using the mean to network level. The linear separability of the aggregated embeddings for each class will be compared.

\subsubsection{Classifying class and sub-class of Peel's quintet}
This section will test the ability of the embedding methods to separate the hidden classes of Peel's quintet. The embeddings generated in the previous experiment will be used, and a multinomial logistic regression will be created for each network to see the accuracy of separating either the nodes into their known classes or into their hidden classes. The logistic regression will use as the independent variables the  two embedding values. In the case of SETSe, these will be the node elevation and the mean node tension. The role of the logistic regression is not so much to be a predictive model but to test the separability of the data; as such, no cross validation will be necessary. The accuracy measure will be accuracy as the classes are balanced. The performance of SETSe will be compared against all other embedding methods.

\subsection{Relationship with assortativity}

The Facebook data will be embedded for all of the 100 universities using the graduation year of the student. Although technically categorical data, graduation year can be treated as continuous and will be done so for speed of calculation. Missing data will not exert a force. The embeddings will be aggregated to network level using the mean elevation and mean node tension. The resulting two-dimensional data will be compared to the assortativity scores of the data to see if there is a relationship between the SETSe embeddings and the network assortativity. 

\subsubsection{Predicting node class in Facebook data}

This tests to see how well SETSe can separate the classes within large and complex networks. Year of graduation will be the embedding class, student type will be the hidden sub-class. The two main classes of student are types 1 and 2, which make up 80\% and 15\% of the dataset, respectively. The meaning of student type is not clear from the original paper, but it appears to be graduate student or alumna. Due to the distribution of student type 2, only 2005 will be used for the hidden sub-class test. Student type is being chosen over the other variables, as dorm, major, minor and high school have so many levels that embedding would be impractical. Gender has two levels only; however, there is almost no assortativity suggesting a complete lack of structure.

Due to the complexity of the data, a $k$ nearest-neighbour approach will be used. This method will label the node as the majority class of the nearest $k$ nodes in SETSe space. The nearest-neighbour model will be compared against graph adjacency voting. Graph adjacency voting finds the majority class amongst all nodes for which the target node shares an edge. The graph adjacency voting will use the full network, but only student types 1 or 2 will count towards the totals. The metrics used to evaluate performance will be accuracy, balanced accuracy, Cohen's kappa and the f1 score. The hidden classes are highly imbalanced in some of the universities, and so the results need to be interpreted with care. The hidden class model accuracy will also be compared against the naive ratio of  type 2 students (the majority class) to all students.

\subsection{Computational details}
\label{sec:comp_deets}
Each simulation used a single core Intel Xeon Gold 2.3 GHz processor with 4 GB of RAM. Code and analysis used R version 4.0, and made extensive use of \texttt{igraph} \cite{csardi_igraph_2006} and \texttt{rSETSe} \url{https://github.com/JonnoB/rSETSe} packages. The non-SETSe embeddings were done using the \texttt{GEM} library \cite{goyal_graph_2018} in Python 3.6.

\section{Results}
\label{sect:results}

\subsection{Peel's quintet}

\begin{figure}
    \centering
    \includegraphics{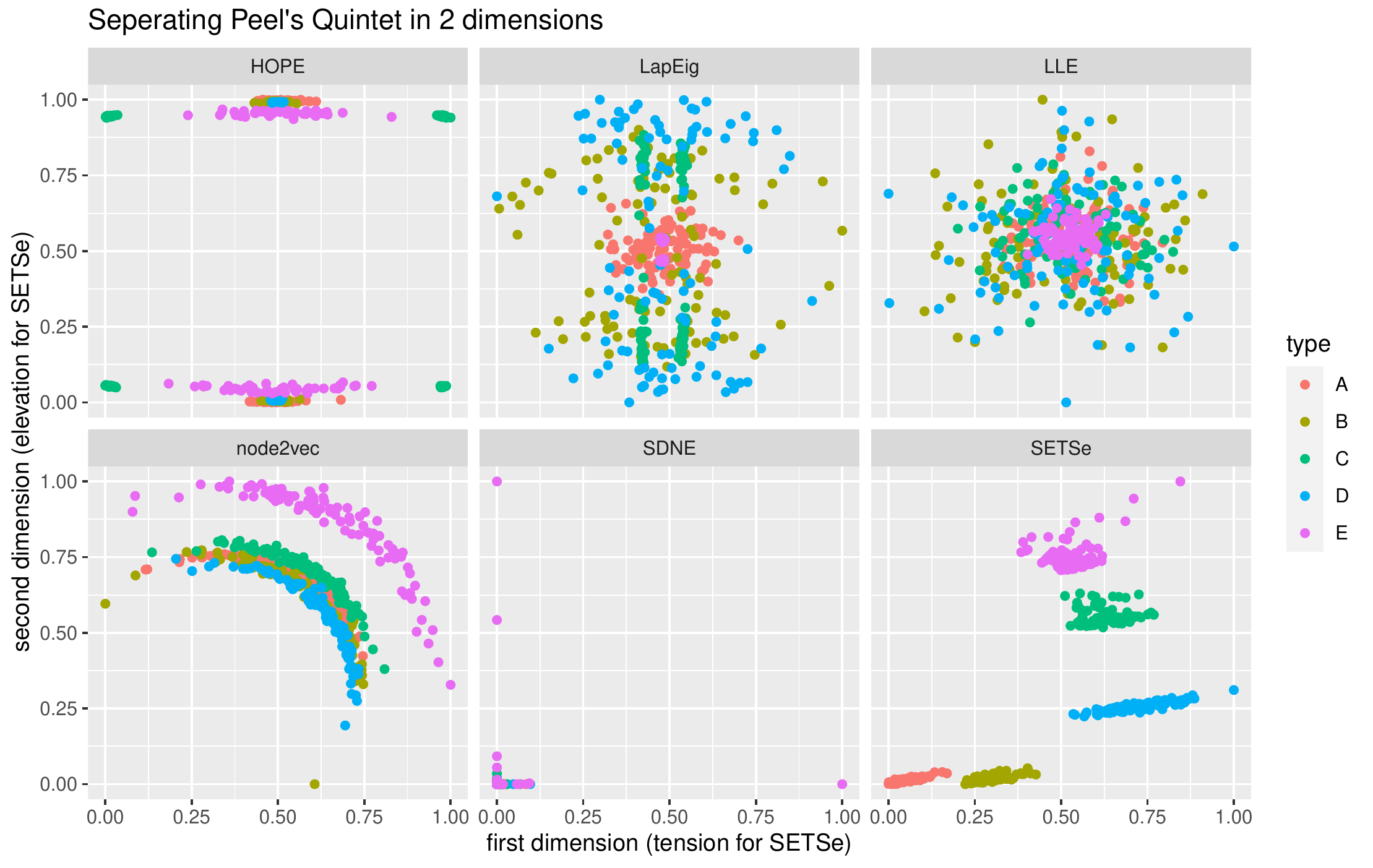}
    \caption{One hundred examples of each of the five  network classes of Peel's quintet introduced by \textcite{peel_multiscale_2018}. The networks are clearly linearly separable in the averaged SETSe space even though all 500 are identical using assortativity.}
    \label{fig:sep_quintet_basic}
\end{figure}

The first test of the SETSe algorithm uses Peel's quintet of networks. The 100 networks of each class are projected into SETSe space then aggregated using the mean absolute elevation and the mean of the node tension. Figure \ref{fig:sep_quintet_basic} shows the results of the six algorithms reducing the networks to two-dimensional space. It is clear that the different connection patterns between the nodes result in distinctive tension elevation patterns within the graph class. This results in the five networks being trivially separable in SETSe space. The other graph embedding algorithms struggle to differentiate the network types, node2vec is the most successful projecting the graph types into more or less concentric quarter rings in two dimensions. The HOPE algorithm also has some success, but like node2vec failed to provide a clear linear separability. The SDNE algorithm was unable to provide successful embeddings; this may be due to the number of embedding dimensions being so low, or due to the structure of the quintet graphs themselves.

\begin{figure}
    \centering
    \includegraphics{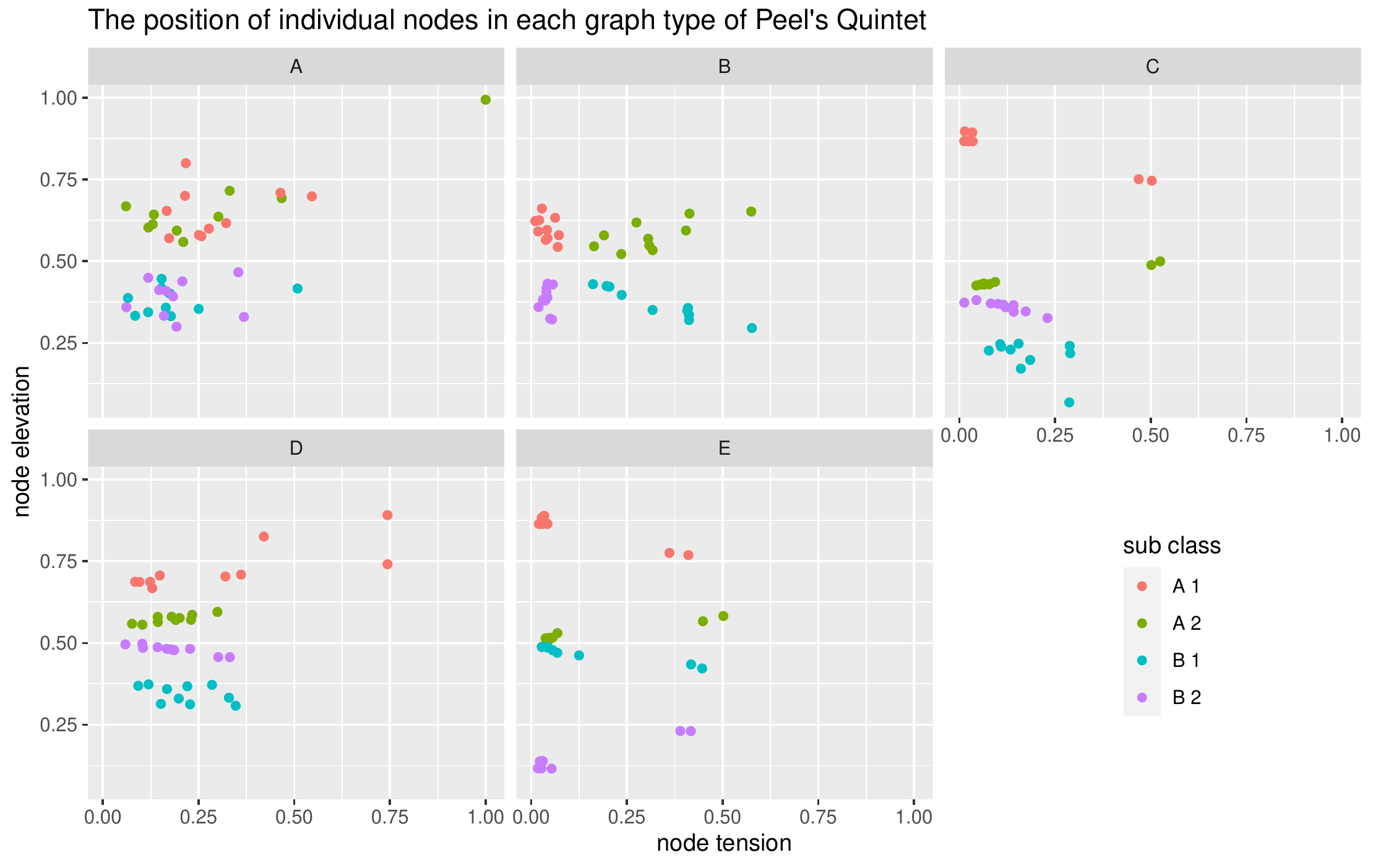}
    \caption{Embeddings of individual nodes; the $x$-axis is node tension and the $y$-axis is node elevation. This representation reveals the hidden structure of the groups in all network types apart from type A.}
    \label{fig:peel_node_embedddings}
\end{figure}

\begin{figure}
    \centering
    \includegraphics{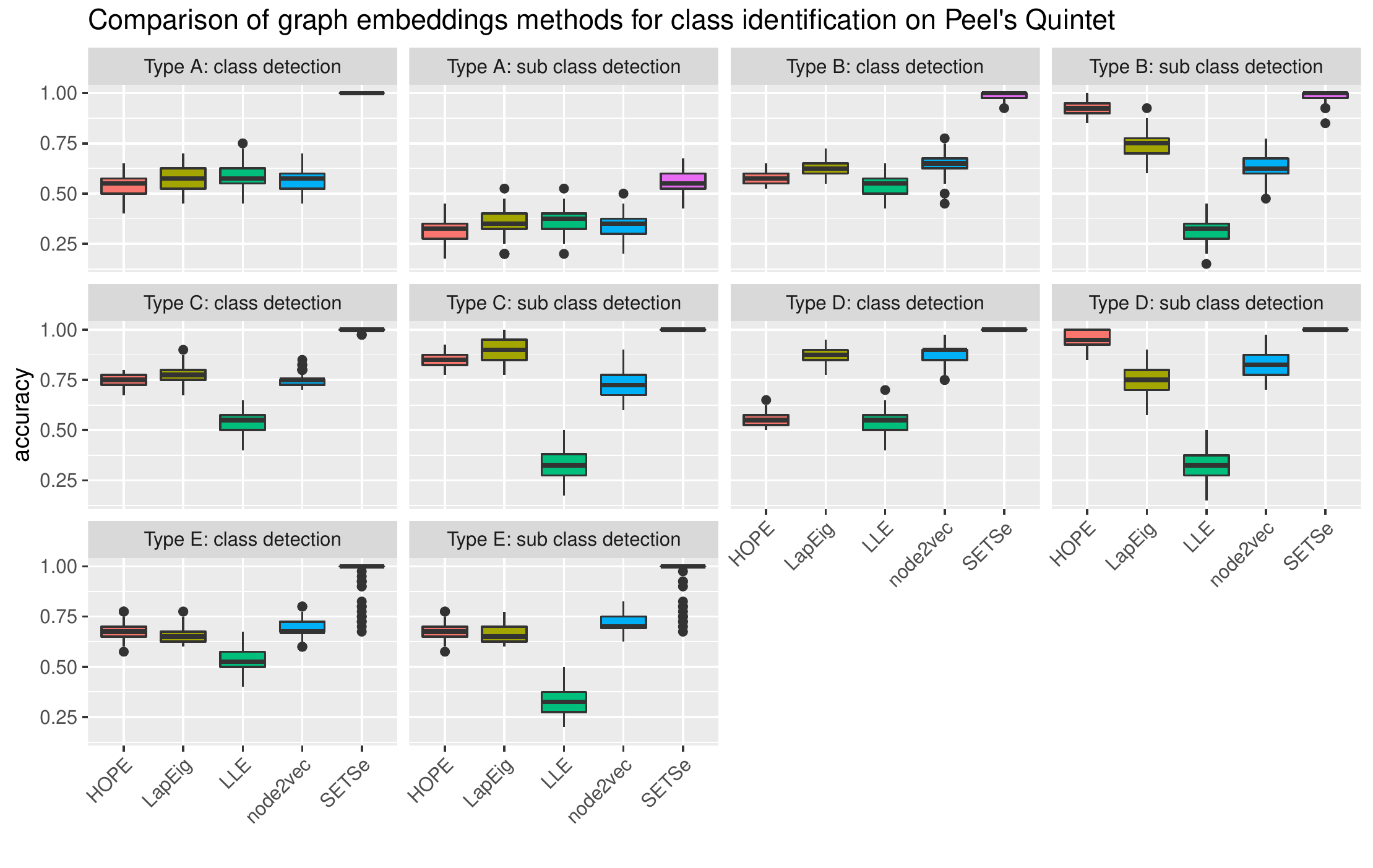}
    \caption{Using the knowledge of the groups, SETSe greatly outperforms the other node embedding techniques, at separating the classes and sub-classes for each type of Peel's quintet. The upper and lower bounds on the boxes describe the 26th and 75th percentiles. The whiskers are 1.5 times the inter-quartile range.}
    \label{fig:peels_clustering}
\end{figure}

Clearly, SETSe is successful at separating the graph types. However, it is also interesting to know whether it can assign the nodes to the correct classes within each graph type. Figure \ref{fig:peel_node_embedddings} shows the SETSe embedding of the nodes in the elevation tension dimensions for an example graph of each type. The nodes are coloured by the hidden sub-class. As can be seen, there are clear patterns in the node placement. Although the sub-classes of type A cannot be distinguished, types C, D and E appear to be linearly separable in the elevation dimension alone. In contrast, type B produces a roughly symmetrical distribution requiring both elevation and strain for separation. The separability is checked for all 500 networks using SETSe and the five other embedding types. The results are shown in figure \ref{fig:peels_clustering}. The figure shows how each embedding technique separates the classes and sub-classes for each type of graph in Peel's quintet. A multinomial logistic regression with two independent variables, reflecting the two dimensions of the embedding, was used to model the accuracy of each class and sub-class within the graphs.

As can be seen from figure \ref{fig:peels_clustering}, SETSe outperforms the other embedding algorithms in every case, again node2vec and HOPE come next in terms of performance. When comparing pure linear separability, SETSe greatly outperforms all other embedding techniques. With the exception of identifying the sub-class of graph type A, SETSe can linearly separate the classes and sub-classes at least 67\% of the time. And it can perfectly linearly separate the sub-classes in four of the ten cases. No other embedding method is close, HOPE can linearly separate all four sub-classes for graph D 32\% of the time, but for most cases, linear separation is not possible on either the known binary class or the four hidden sub-classes. The SDNE algorithm was not included in figure \ref{fig:peels_clustering} due to poor performance. When the other methods are allowed to embed the 20 node graphs in eight dimensions, their performance increases substantially to a level comparable with SETSe.
As an additional comparison, three community detection algorithms were also compared: Fast Greedy \cite{clauset_finding_2004} , Walktrap \cite{pons_computing_2006} and Louvain \cite{blondel_fast_2008}. These algorithms were not successful at distinguishing between the classes or sub-classes.

\subsection{Analysing Facebook data using SETSe}

\begin{figure}
    \centering
    \includegraphics{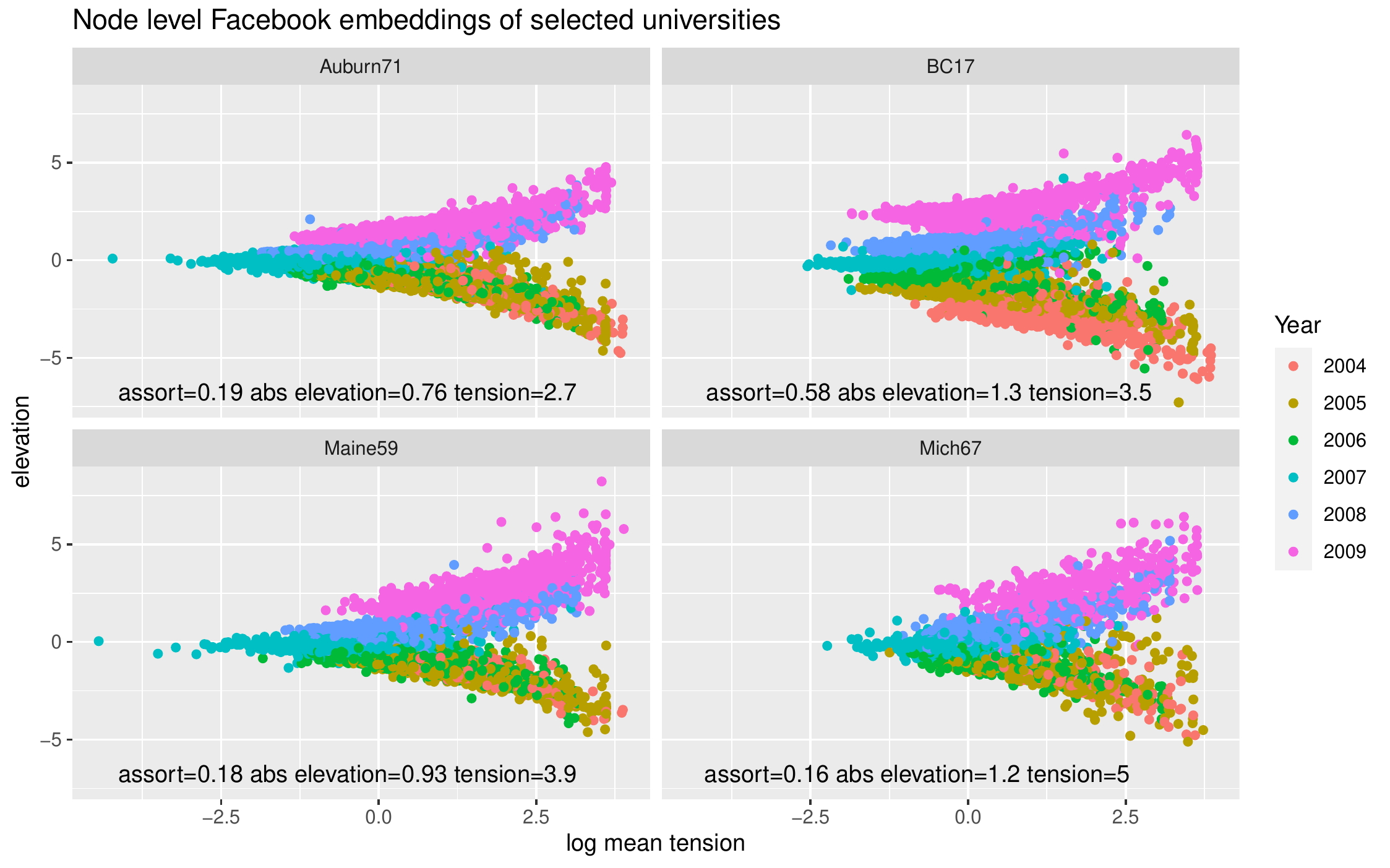}
    \caption{The node level embedding shows that the tension and elevation distribution of nodes by type are distinct and related to the overall network tension and elevation; this is not possible to explore with assortativity.}
    \label{fig:uni_embeddings}
\end{figure}

The Facebook 100 dataset was embedded into SETSe space using the variable graduation year. The results for four different universities are shown in figure \ref{fig:uni_embeddings}. For ease of viewing, the $x$-axis is presented on a log scale. Figure \ref{fig:uni_embeddings} shows three universities with low assortativity (Auburn, Maine and Michigan), meaning that there is a high degree of mixing between years, and one university with a high assortativity (BC), meaning students tend to associate within years. Although it is not possible to explain why the universities have these differences (\textcite{peel_multiscale_2018} suggest it is due to housing allocation policy), it is possible to analyse the relation between assortativity and SETSe. It is clear from the figure that the years separate to roughly their own tension elevation band within the data. This is what we would hope to see given that the embedding uses force based on the graduation year. The overall shape of the data is a funnel with the `nose' at the low-tension end and the funnel at the higher-tension end. It is clear from the plot that the younger years (those graduating in 2008 and 2009) are more separate than the older years (graduation in 2004, 2005 and 2006). This is because they have had less time to form cross-year bonds, and so are more assortative.
The nose is created because the nodes that are most central within their year experience the least tension.

\begin{figure}
    \centering
    \includegraphics{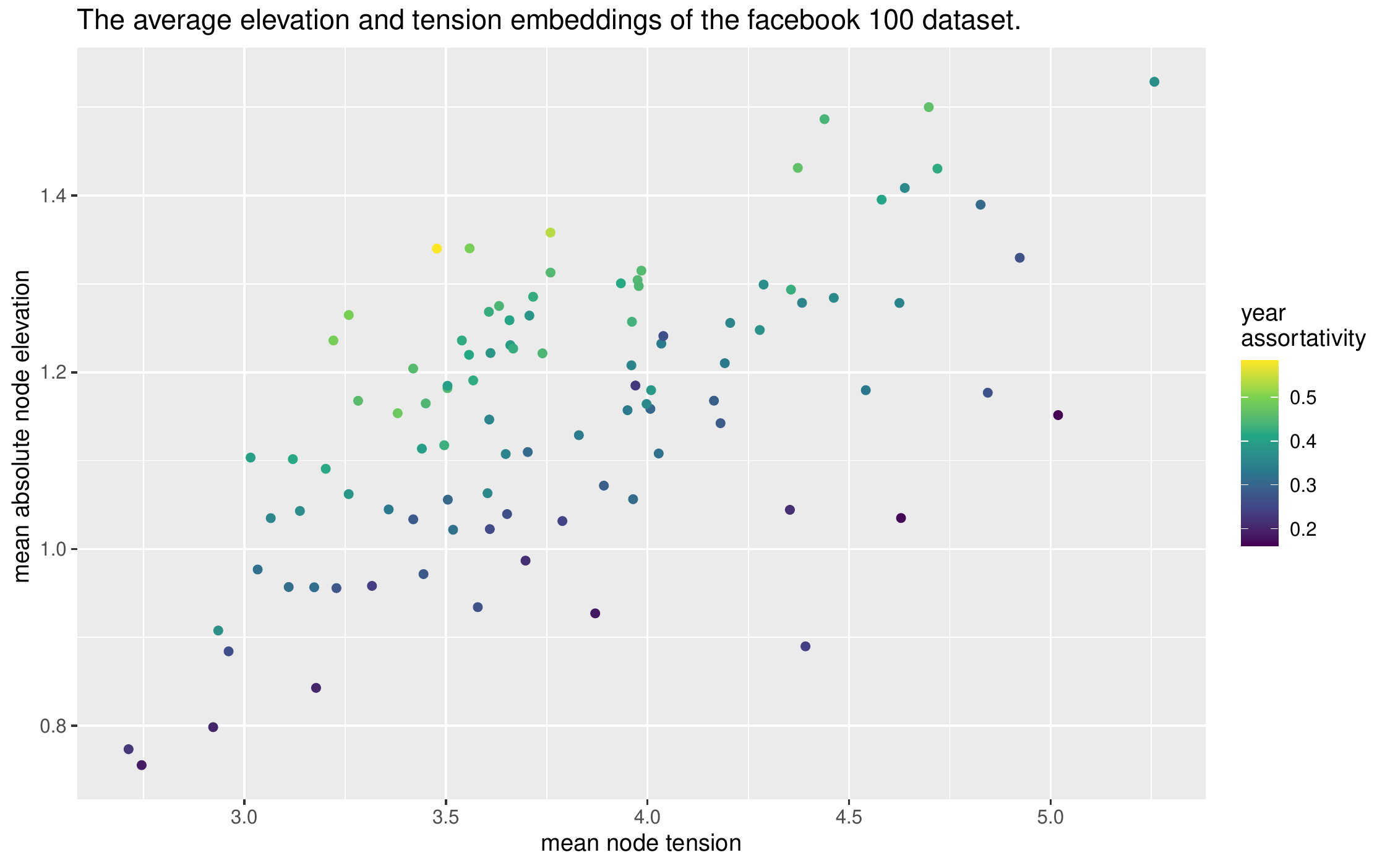}
    \caption{There is a clear relationship between the SETSe embeddings and year assortativity; however, the pattern is dependent on elevation and tension. SETSe finds large differences between networks with very similar assortativity.}
    \label{fig:facebook_100_elev_tens}
\end{figure}

The three low assortativity universities have very different tension and elevation scores. It can be seen that higher tension appears to create a fuzzier, less clearly defined groups within each year. Low elevation creates a single `nose' for the whole cone, but higher elevation starts separating out, forming a nose for each year.
It is impractical to understand the differences of the universities by looking at the scatter plots of thousands of students across 100 universities. The elevation and node tension are thus aggregated at university level, and all 100 universities are plotted in Figure \ref{fig:facebook_100_elev_tens}. The points are coloured by assortativity. It can be seen that while there is a positive relationship between tension and elevation, there is, in fact, a negative relationship between the tension and elevation dimensions and the value of assortativity. Analysing the distribution of the tension, elevation and assortativity scores for all 100 universities, it is found that the data are normally distributed. Creating a linear regression on assortativity using tension and elevation as independent variables provides an $R^2=0.82$, where the coefficients are significant to $p<0.001$. Creating linear models using either tension or elevation provides $R^2=-0.006$ and 0.429, respectively.  

Next, the ability of the embeddings to be able to predict the $k$ nearest-neighbour nodes is tested against a baseline of graph adjacent voting. The model is an effective predictor of year with high values of kappa (59\% when $k$ is 9), and balanced accuracy (77\% when $k$ is 9) indicating that the model predicts above the naive baseline value. However, the graph adjacency model outperforms the knn using SETSe by between 7\% to 10\% in terms of accuracy; kappa and f1 score and up to 15\% on balanced accuracy. It should be noted that high levels of performance are expected as the model is predicting on the data it was embedded with. Despite this caveat and the poor performance against the adjacency voting model, the results show that the embeddings produced are meaningful, even on large and complex networks, although the actual graph structure is lost.

\begin{figure}
    \centering
    \includegraphics{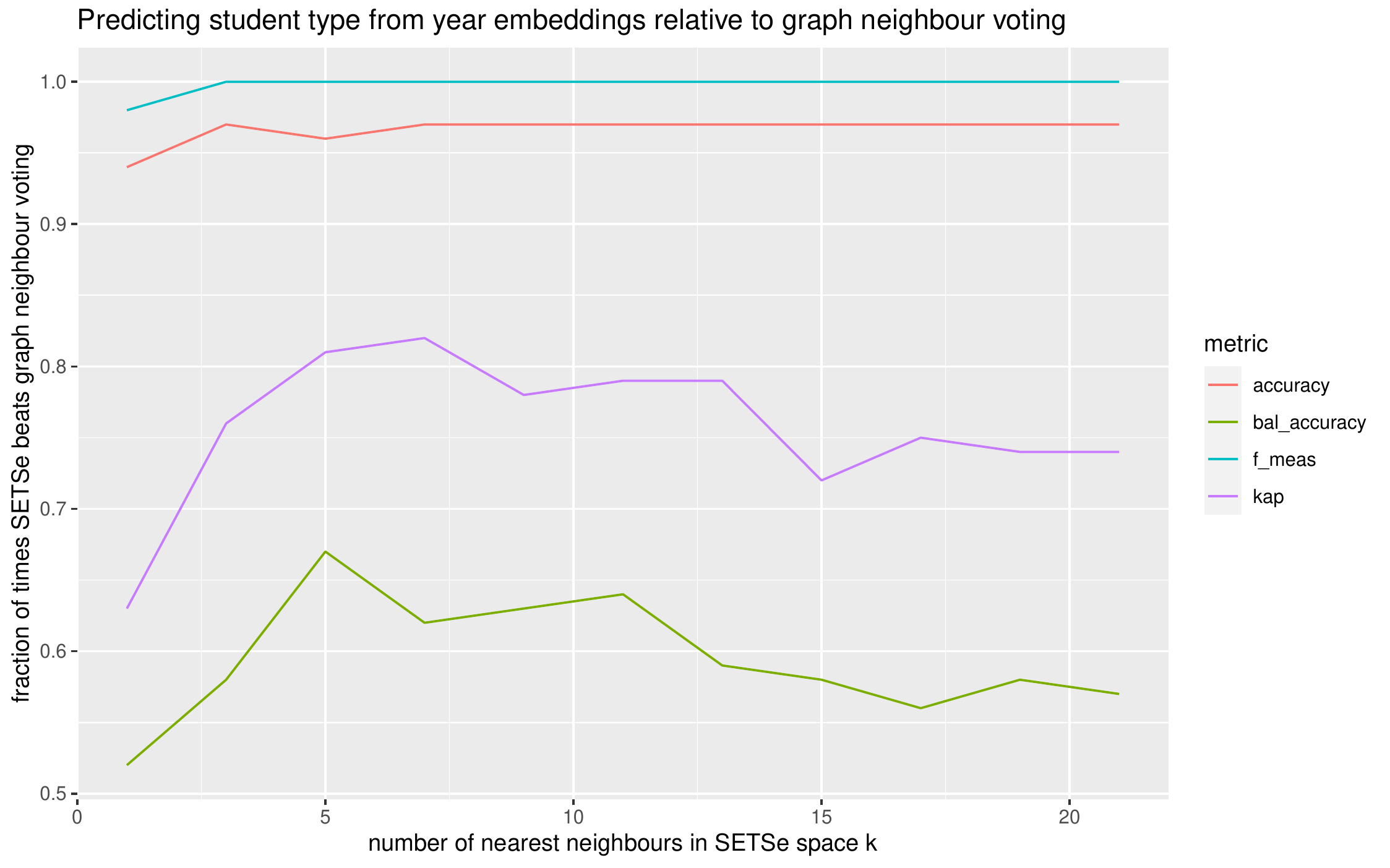}
    \caption{Using $k$ nearest neighbours to predict the type of student performs better. This is despite the graph embeddings being relative to year not student type. SETSe beats the baseline more than 50\% of the time for all values of $k$. Note for ease of viewing, the $y$-axis starts at 0.5.}
    \label{fig:facebook_knn_vs_graph}
\end{figure}

Figure \ref{fig:peels_clustering} showed that SETSe could uncover the hidden structure of the network; however, the question is can SETSe do that on a complex real-world dataset? Using the same year embeddings for the Facebook 100 dataset, the $k$ nearest neighbours were used to predict student type. The results show that the accuracy of the student type model averaged across all 100 universities, outperforms the naive rate of student type 2 over student type 1 at all values of $k$. The model performance is also quite stable for all values of $k$, accuracy is around 71\%, balanced accuracy is 55\%, kappa is very low at 12\% and the f1 score is around 79\%. The low balanced accuracy is not significantly higher than 0.5, and the low kappa score signifies that the results could simply be due to chance. However, as shown in figure \ref{fig:facebook_knn_vs_graph}, the model comprehensively beats the nearest-neighbour voting method in all metrics for almost all values of $k$. Note that the f1 score is not reliable in this case as there are a significant number of occasions where type 1 students are not predicted at all. This means that the f1 score is dependent on the class labelling; balanced accuracy and Cohen's kappa avoid this issue.

\subsection{Complexity}
\label{sect_complexity}
\begin{figure}
    \centering
    \includegraphics{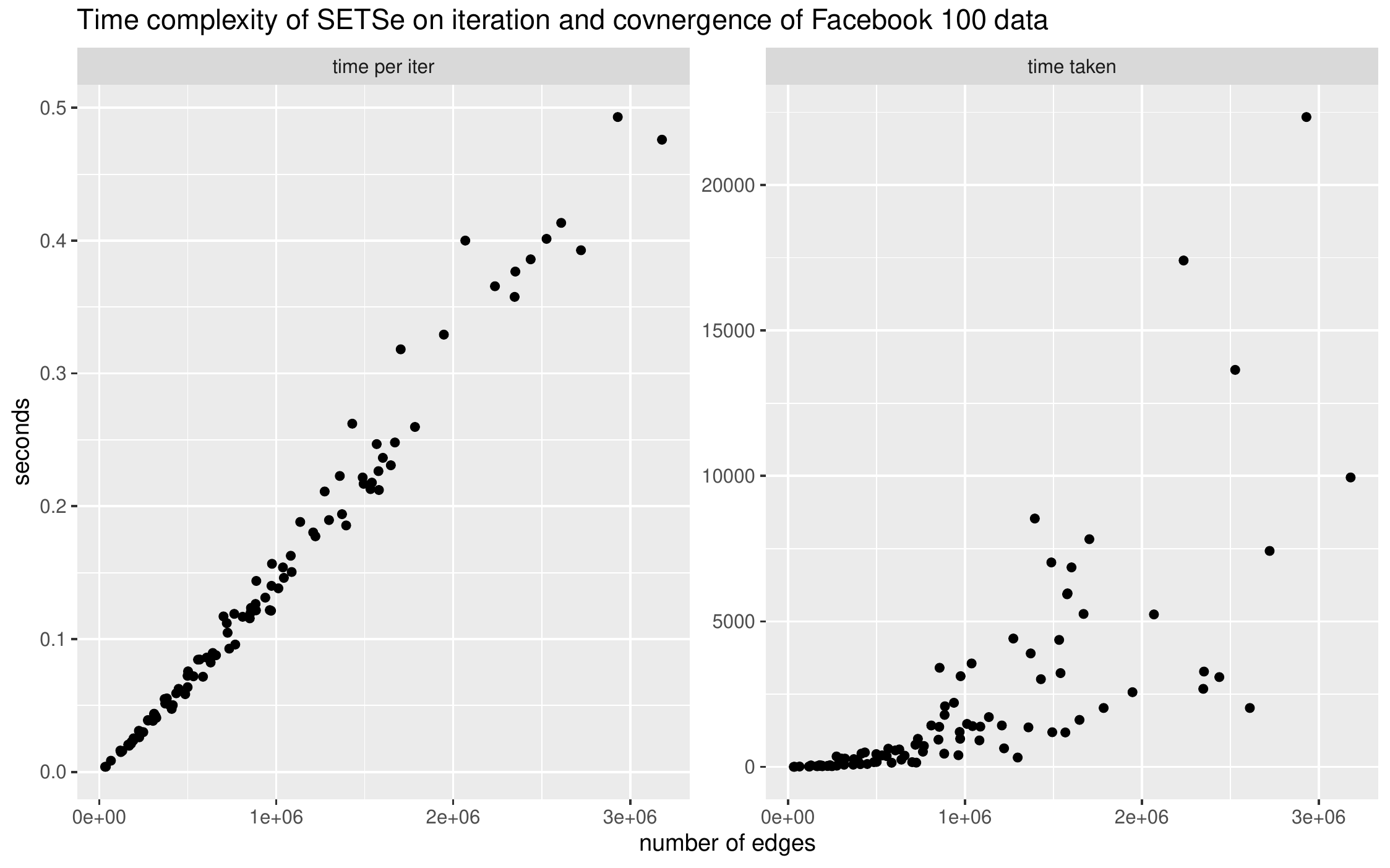}
    \caption{The per-iteration complexity is linear; however, the complexity of the total time to convergence is closer to quadratic.}
    \label{fig:complexity}
\end{figure}

The time taken to embed the Facebook graph is plotted in Figure \ref{fig:complexity}. The left panel of the figure shows that the iteration time complexity is almost perfectly linear ($\mathcal{O}(\left | \textrm{E} \right |)$). The time taken to reach convergence is shown in the right panel. The convergence complexity shows heteroscedasticity and is closer to running in quadratic time ($\mathcal{O}(\left | \textrm{E} \right |^2)$). This is in keeping with other spring system models that are also ($\mathcal{O}(n^2)$), but is slower than LLE, Laplacian, Eigenmaps and HOPE, which run in ($\mathcal{O}(\left | \textrm{E} \right |)$), as well as node2vec and SDNE, which run in ($\mathcal{O}(\left | \textrm{V} \right |)$), for single attribute networks. It should be noted that time to convergence is highly dependent on the network topology; see the Appendix on the bi-connected component method. On the system described in section \ref{sec:comp_deets} a network of 40,000 nodes and 1.5 million edges uses about 1.9 GB of RAM, to converge and 3.9 hours.

\section{Discussion}

SETSe acts as a hybrid between the advanced techniques of the graph embedders used for analysis and the intuitive simplicity of the spring embedders used for graph drawing. It does this by projecting the network onto an $n+1$-dimensional manifold, using node attribute as a force. This is different from traditional spring embedders where all nodes exert an equal force \cite{kamada_algorithm_1989, fruchterman_graph_1991, eades_heuristic_1984}. 
As such, unlike the graph embedding algorithms that were used to benchmark performance in this paper, SETSe cannot be said to `learn' the properties of the network. Instead, similar to the other spring embedders, SETSe finds an equilibrium position wherein all forces are balanced. SETSe also distinguishes itself from the graph embedders and spring embedders by being entirely deterministic. The SETSe algorithm is fast within each iteration, running in $\mathcal{O} ( \left | \textbf{E} \right |)$ linear time. However, the time to convergence is not linear and appears to be closer to $\mathcal{O} (\left | \textbf{E} \right |^2)$, which is slower than most machine learning embedders.

When separating Peel's quintet, only SETSe managed to find a successful two-dimensional representation of the networks. Two points should be made on this, as the output of the other graph embeddings produces a two-dimensional representation of each node. SETSe produces a one-dimensional representation of each node and a two-dimensional representation of each edge (although strain and tension are identical in this case). Another point is that the other graph embedding algorithms can embed the graph in any number of dimensions; typically the graph would be embedded in higher-dimensional space then visualised in two dimensions by embedding the nodes a second time using some other data reduction method \cite{van_der_maaten_visualizing_2008, pearson_liii_1901}. These two points illustrate the fundamentally different approach to embedding that SETSe takes as it is able to natively project high-dimensional data in meaningful low-dimensional space without needing a secondary embedding method. The explicit edge embedding also allows for flexibility when it comes to projection choices, which is something other embedding methods lack. The goal of typical graph embedders is to minimise the distance between nodes according to some measure of similarity. In contrast, SETSe does not try to optimise the meaning in the data; instead, it maps the attributes and edge weights of the network to a new space (the manifold), and in doing so reveals properties of the network.

The work of \textcite{goyal_graph_2018} suggests that at least some of the other graph embedders would have been able to linearly separate the sub-classes of Peel's quintet more successfully if more dimensions were used. However, this leads to new problems such as how many dimensions should be used? What dimension reduction algorithm should then be chosen to reduce the dimensions again for plotting purposes, and how can the results be interpreted? In the case of using the output of the embedding for a model, dimensional parsimony is paramount. Being able to model the dependent variable in two dimensions is much more desirable than getting the same results with 16 dimensions. This is not to say that SETSe is always better than the other methods tested here. In particular, for applications where node similarity is the most important feature, or a large number of dimensions is advantageous, it is likely to be outperformed. Also, unlike most machine learning graph embedders, SETSe is unable to perform link prediction. Being bound by physics also means SETSe cannot be easily tuned to target specific goals. However, while most graph embedders are designed to express a similarity chosen by the designer and parametrised by the user, SETSe in contrast is broadly goal agnostic. Instead, it simply allows the graph to express in a different way what was already there and provides insight into the underlying data in the process. In some cases, SETSe could be used in the preprocessing stage of the more sophisticated graph embedders, providing edge and attribute data to support similarity optimisation.

\section{Conclusions}

The SETSe algorithm is an unsupervised graph embedding method that uses node and edge attributes to project a graph into three separate latent vector spaces. The node vector space is pairwise Euclidean for connected nodes. Thus the projection of data into SETSe space can be considered a manifold of dimension $n+1$, where $n$ is the number of variables, and the final dimension is the graph space defining the minimum node distance. The vector space provides insight into the original data and can reveal the structure that was not available when the embedding was produced. 

This paper showed that SETSe outperforms several popular graph embedding algorithms, on tasks of network classification and node classification. SETSe also provides distinctions between networks that appear to be similar or identical when using the popular assortativity metric. 

SETSe is a physical system whose equilibrium state allows previously obscured patterns in the data to be expressed. And although it could be classed as an unsupervised learning algorithm, SETSe does not learn any properties of the system or attempt to optimise similarity. As such, it can be considered both as an auxiliary of, and a counterweight to, machine learning techniques. This is important because although the value of machine learning in current research progress cannot be overstated, it is not the answer to everything. Besides, often the more sophisticated a technique, the more subjective choices are required, both in development and parametrisation.

Although spring embedders lost popularity with the rise of machine learning, SETSe shows that there is still a role for unsupervised physics models in modern data analysis. The spring has bounced back.

\section{Additional materials}
The Appendix contains details on the auto-SETSe and bi-connected SETSe algorithms used to make network convergence easier.

In addition an R package has been created, \texttt{rSETSe}, which can be used to create SETSe embeddings. The package allows easy embedding of graphs on networks of tens of thousands of nodes and over a million edges on a normal laptop. The package can be installed from \url{https://github.com/JonnoB/rSETSe}.

\section*{Acknowledgements}
I would like to thank Connor Galbraith and Patrick De Mars for their thoughtful and patient advice at all stages of this project; Dr Ellen Webborn for her insightful and thorough feedback on the manuscript; and my supervisors Dr Elsa Arcaute and Dr Aidan O'Sullivan for giving me the space to pursue this idea.
I acknowledge  use of the UCL Myriad High Performance Computing Facility (Myriad@UCL), and associated support services, in the completion of this work.

\printbibliography

\end{document}


\maketitle

This appendix describes the methods used to find the converged positions of the SETSe algorithm substantially faster than using guess-work. The two main methods are auto-convergence, which finds optimal parameters for a given network, and the bi-connected method, which breaks the network into smaller pieces and solves for each piece then re-assembles.

\section{Auto-convergence}

\begin{figure}
    \centering
    \includegraphics{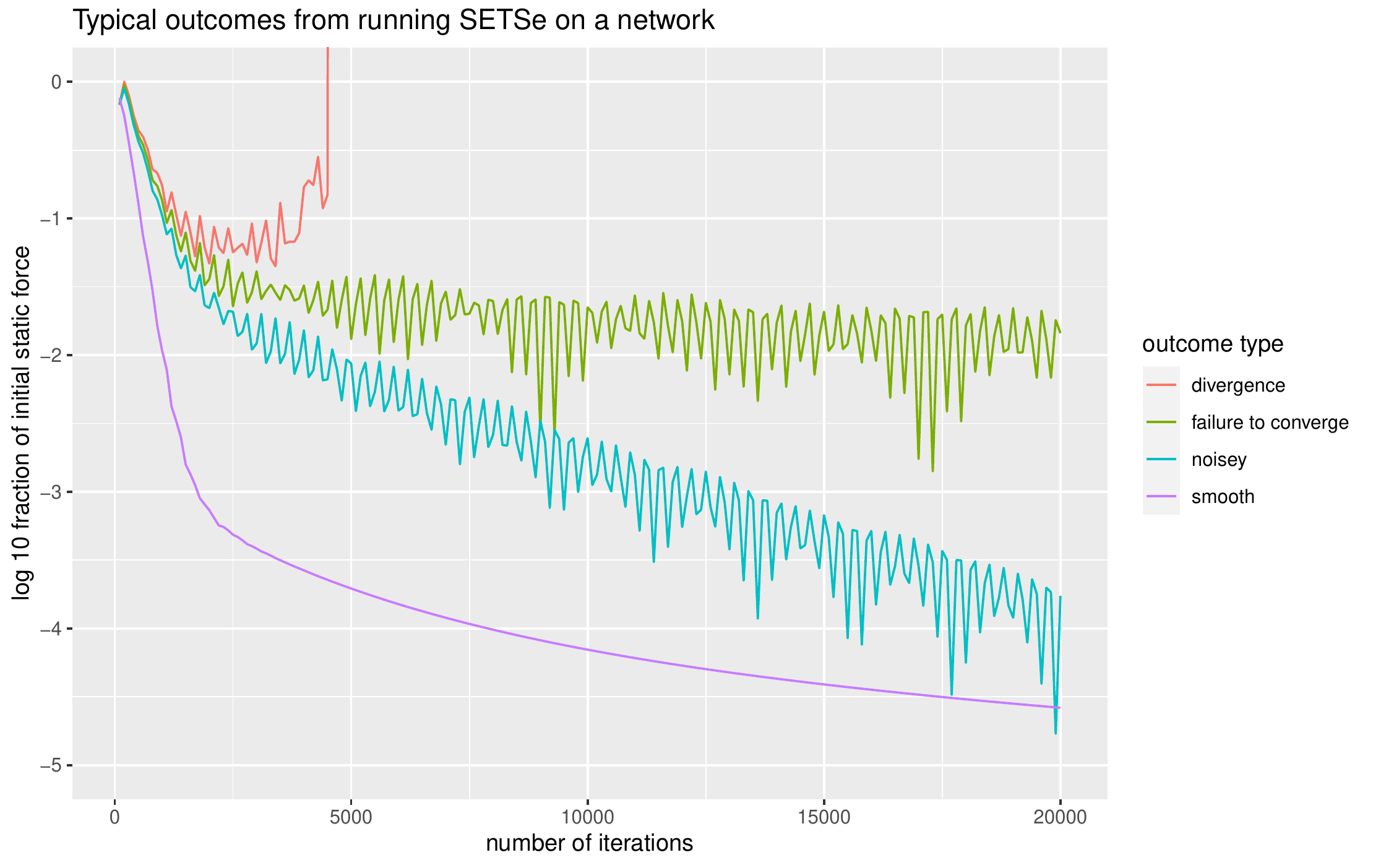}
    \caption{Different outcomes from using SETSe on the same dataset: changes caused by different values of the coefficient of drag. }
    \label{fig:outcomes}
\end{figure}

The SETSe algorithm must be correctly parametrised or the system will not converge. The outcome of the convergence of a SETS embedding is mostly controlled by the relationship between the time step and the coefficient of drag. The four outcomes from SETSe are shown in figure \ref{fig:outcomes}. Figure \ref{fig:outcomes} also shows the static force at each iteration of convergence for the Caltech36 Facebook network. The convergence parameters are identical in all cases, apart from the coefficient of drag $c$. The four outcomes are divergence ($c= 0.65$), failure to converge ($c = 0.7$), noisy convergence ($c = 0.75$) and smooth convergence ($c = 1.2$). Divergence occurs as a result of the  discrete time step; if the time step is too large relative to the drag, the system gains energy, causing the kinetic energy and spring energy of the nodes to explode. Failure to converge occurs when the system is in a stasis where it can neither lose energy nor diverge. With noisy convergence, the system loses energy on average, but the system goes through erratic jumps caused by force shocks propagating through the network (discussion of shock propagation is beyond the scope of this paper). As can be seen from the figure, smooth convergence is the most desirable as it reaches the converged state most quickly. It should be noted that in figure \ref{fig:outcomes} the outcomes `noisy convergence' and `failure to converge' have two waveforms superimposed on each other. This appears to be due to force shock waves moving back and forth across the network.

\begin{figure}
    \centering
    \includegraphics{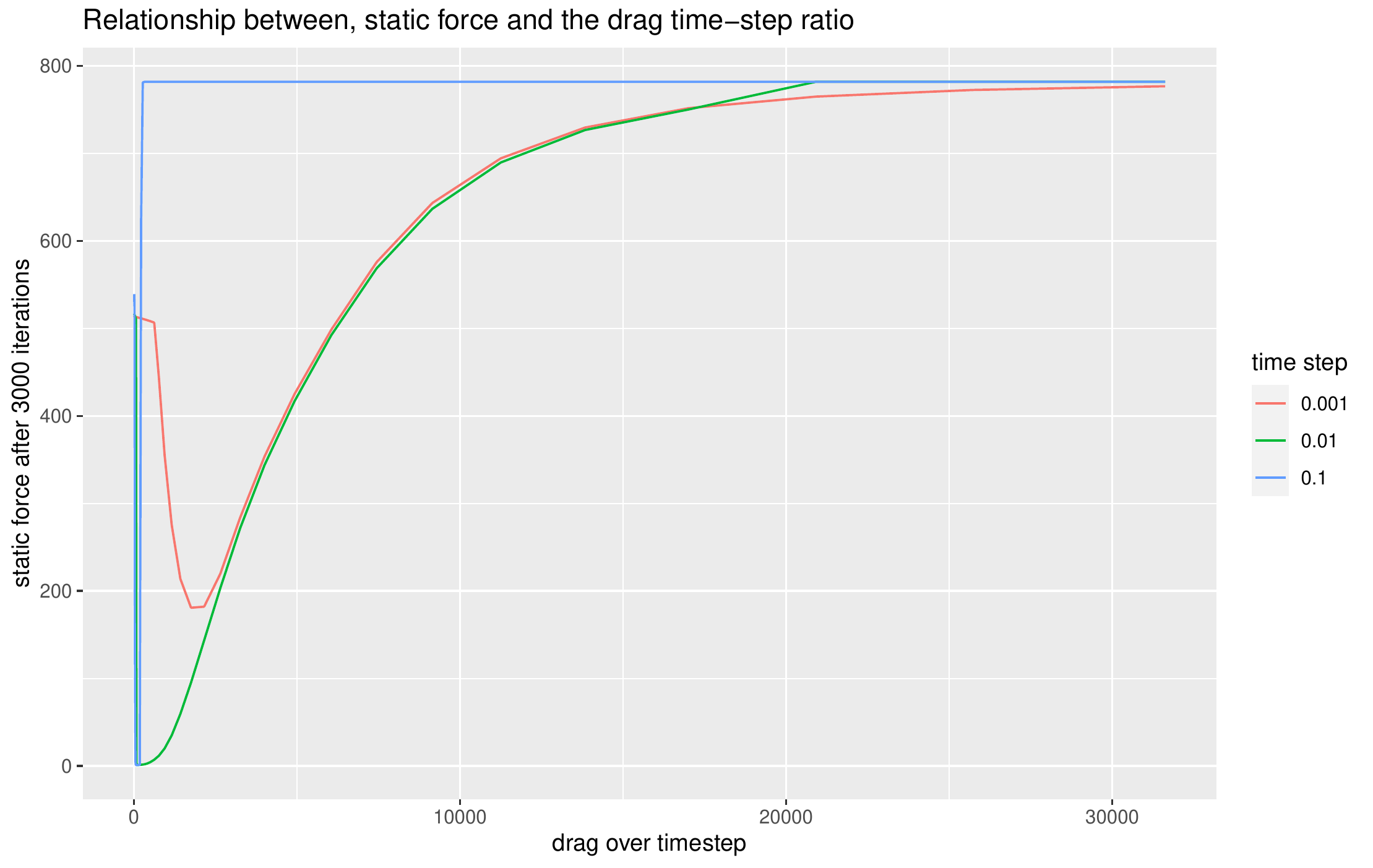}
    \caption{The static force at after a set number of iterations is convex relative to the ratio of drag to time-step size.}
    \label{fig:static_force_vs_drag_tstep}
\end{figure}

Figure \ref{fig:static_force_vs_drag_tstep} shows the final static force versus the coefficient of drag for three time-step values. The axes of the plot are the static force after 3000 iterations and the ratio of the drag coefficient to the time step. The figure shows that the static force is a convex function of the ratio. This convex shape allows a search to be used to find a parametrization close to the optimum, and is the basis for the auto-SETSe function. 

It should be noted that figure \ref{fig:static_force_vs_drag_tstep} shows that smaller values of $\Delta t$ give a larger search space to find optimal conditions.

The auto-SETSe algorithm is shown in algorithm \ref{algo:autoSETSe}, which takes advantage of the convex shape of the static force under different drag values. 
Auto-SETSe uses the same parameters as the SETSe algorithm; in addition, it uses a set of hyper-parameters used to tune the value $s$. The auto-SETSe algorithm runs in a while loop with several termination conditions, the first of which is that the current iteration $p$ does not exceed the maximum allowed number of iterations $p_\textrm{max}$. This prevents the loop running infinitely.

Simulations are terminated early when the static force exceeds the total absolute force produced by the nodes because, under optimal parametrization, this does not occur. The simulations are run using a much shorter number of iterations than during normal SETSe. The algorithm searches the log 10 space between a user defined highest a lowest possible drag values. If none of the drag values get static force values lower than the initial value then the time step is reduced and the process repeated. Once the simulation has found a value of drag and timestep combination that smoothly converges a binary search is performed to find the optimum value. With the optimum value identified, the simulation is run using a larger number of iterations (of the order of 10 times larger), this final stage adjusts the time-step if the process becomes noisey which can occur when the static force becomes small compared to its initial value. This approach using early termination for sub-optimal parameters and much shorter simulation times makes finding parameters that are fast and delivering high-quality results straightforward.

The static force after the final iteration of the SETSe algorithm is $\eta = \sum \left | f_{i} \right |$, where $\eta$ is the sum of the absolute value of the static force experienced by all nodes in the network. The value $\eta_0$ is the static force and initialisation. $\lambda$ is the convergence tolerance; once $\eta$ has fallen below this level, the system is said to have converged and therefore further parameter search is not required. Auto-SETSe is said to have found optimum parameters when the value of $\eta$ stabilises. The system is said to be stable when the previous best value of $b$ ($ b_\textrm{min}$), and the current value are within  $\phi$ of each other. This is shown on line 9 as $b_p$, if $\eta_p<\eta_0$ and is twice the tolerance otherwise, shown on line 12. If $p \geq p_\textrm{max}$ or $b_p < \phi$ but the system has not converged, meaning $\eta_p>  \lambda$, then SETSe is run again using a substantially larger number of iterations using the best value of $c$ found by the auto-SETSe process. 

Whilst this process is not guaranteed to result in a converged SETSe, it provides satisfactory embeddings in most cases.

\begin{algorithm}
\SetAlgoLined
\KwResult{Graph $\mathcal{G}$ embedded on best value of $c$ for given $\Delta t$}
 $p = 1$\;
 $s_p= 0$\;
 \While{$(\eta_p>  \lambda) \; \& \; (b_p > \phi) \; \& \; p < p_\textrm{max}$}{
 $c_p = \frac{\Delta t}{10^{s_p}}$\;
     Run SETSe using $c_p$\;
   Store $\eta_p$ the output of SETSe along with $c_p$\;
  \eIf{$\eta_p< \eta_0$}{
  Set $s_p+1$ using the next value produced by a binary search process on the stored values of $s$\;
  $b_p = \frac{ \eta_p - \eta_\textrm{min} }{ \eta_\textrm{min} }$
   }{
   $c_p= 10^{\textrm{log}_{10}(c_p)+1}$\;
   \If{$c_p>$max drag }{
   $c_p=$ drag min\;
   $\Delta t = \Delta t \cdot$ time step shrinker;
   }
   $s_{p+1}= \textrm{log}_{10}(\frac{\Delta t}{c_p})$\;
   $b_p = 2b_\textrm{min}$
  }
  $p = p +1$\;
 }
 \If{$\eta_p> \lambda$}{
 
 Run SETSe using best value of $c$ and a larger number of iterations
 
 }
 
 \caption{The auto-SETSe algorithm.}
 \label{algo:autoSETSe}
\end{algorithm}

\section{The bi-connected component solution}

Although auto-SETSe is very valuable, there remain convergence issues even on ideal parameters. Post-convergence analysis of the Caltech36 network shows the 31 nodes (4\% of total graph), which have only a single neighbour, make up 31\% of the static force.
This is intuitive as nodes on the periphery of the network or nodes that connect two parts of the network together can be subject to large and unpredictable changes in experienced force, much like a double pendulum. This means that a slight change in position due to group nodes moving can produce substantial static force on these peripheral or joining nodes.
The most effective way to solve this is to break the network up into its bi-connected components.
A bi-connected component is a sub-graph that will form a separate component with the removal of a single node, called an articulation node. Such nodes act like bridges between groups of nodes and so are prone to experiencing sudden shifts in force magnitude and direction. When a network is decomposed into its bi-connected components, it is called a block-cut tree. An example of a network with four bi-connected components is shown in figure \ref{fig:bi_conn}. In the figure, articulation nodes are named $a_i$, where the suffix $i$ indicates the index of the articulation node. 

\begin{figure}
    \centering
       \begin{circuitikz}
    [scale=.8,auto=left,every node/.style={circle,fill=blue!20}]
      \draw (0,0)
      to[R=] (2,-2) 
      to[R=] (4,-4) 
      to[R=] (6,-4)
      to[R=] (6,-2)
      to[R=] (4,-4);
      \draw (4,0)
      to[R=] (2,-2);
      \draw (4,0)
      to[R=] (6,-2);

       \draw (0,0)
      to[R=] (2,2)
      to[R=] (0,4)
      to[R=] (-2,2)
      to[R=] (0,0);
      
      \draw (4,0)
      to[R=] (6,2)
      to[R=] (6,0)
      to[R=] (4,0);


      \node (n12) at (0,4) {$n_1$};
      \node (n13) at (2,2) {$n_2$};
      \node (n14) at (-2,2) {$n_3$};

      \node (n1) at (0,0) {$a_1$};
  
      \node (n2) at (4,0) {$a_3$};
      \node (n3) at (2,-2) {$a_2$};
      \node (n4) at (4,-4) {$n_5$};
      \node (n5) at (6,-4) {$n_6$};
      \node (n6) at (6,-2) {$n_4$};
      
     \node (n32) at (6,2) {$n_7$};
     \node (n33) at (6,0) {$n_8$};

       \draw[thick,blue] \convexpath{n1, n14, n12, n13}{0.7cm};
       \draw[thick,red] \convexpath{n1, n3}{0.7cm};
       \draw[thick,green] \convexpath{n3, n2, n6, n5, n4}{0.7cm};
        \draw[thick,yellow] \convexpath{n2,n32,n33}{0.7cm};

    \end{circuitikz}
    \caption{An example network with the bi-connected components highlighted. Articulation nodes are named as $a_1$, $a_2$ and $a_3$; if removed, these nodes cause the network to break into two components. This network includes four bi-connected components.}
    \label{fig:bi_conn}
\end{figure} 

Because the force in SETSe is transferred through the edges, the entirety of the information about the sub-graph is transferred through a single edge. This means that the sub-graph can be simplified to the node that connects it to the rest of the graph (also called the articulation point). The force of the node representing the simplified network is the opposite of the force in the remaining network. By decomposing the network into a block-cut tree made up of all bi-connected components, the degrees of freedom of the network as a whole are greatly reduced. The reduction of the degrees of freedom of the network allows faster convergence of the network as a whole. The bi-connected method functions in a similar way to the Barnes--Hut algorithm \cite{barnes_hierarchical_1986}, which has been applied to reduce the complexity of simulations in astrophysics and has also been applied to some spring embedders \cite{quigley_fade:_2001}. Applying the bi-connected algorithm \cite{hopcroft_algorithm_1973} is efficient as it runs in linear time. 
The restricted movement and constant force direction allow the bi-connected method to find the final position on the manifold in a way that is not possible for the double pendulum mentioned earlier. The simplification works because only the net force of the other bi-connected components is considered (which is static), and the dynamics are not considered during the solution.

The basic outline of the bi-connected component method is shown in algorithm \ref{algo:biconnSETSe}. First, graph $\mathcal{G}$ is decomposed into its bi-connected components. Each component is balanced such that the net force from the connecting component is simplified into the articulation node for that bi-connected component. Then, each bi-connected component is projected into SETSe space. This creates a list of relative node elevations. The function then calls a sub-routine (algorithm \ref{algo:abselevation}), which combines all the relative elevations into a single absolute elevation.

A bi-connected component can be connected to many other bi-connected components, and an articulation node can be connected to many bi-connected components. This means the conversion from relative to absolute elevation values must be treated carefully. The algorithm that does this is shown in algorithm \ref{algo:abselevation}. This algorithm takes the largest bi-connected component to be the origin point of the network and incrementally adds additional bi-connected components to the origin. This means that any bi-connected component being added will be added directly on to the main network not to another relative block. The function takes a list of relative graphs $\textbf{G}_\textrm{vect}$, then initialises an empty graph $\mathcal{G}_\textrm{abs}$ and an empty vector of articulation points $\textrm{Art}_\textrm{vect}$. It adds the origin block to the empty graph. It then adds all the articulation points to the articulation vector. It then takes all the relative graphs that contain the $n\textrm{th}$ articulation points and converts the relative elevation to absolute by subtracting the elevation of the articulation point on the target bi-connected component $G_x$ from all nodes on $G_x$ and adding the elevation of the same articulation point on $G_\textrm{abs}$ to all nodes on $G_x$. The bi-connected component $G_x$ is then added to a temporary graph $\mathcal{G}_\textrm{new}$ until all bi-connected components connected to the $n\textrm{th}$ articulation point on $\textrm{Art}_\textrm{vect}$ have been converted to absolute values. This process continues until all relative graphs have been combined into a single absolute graph.

\begin{algorithm}
\SetAlgoLined
\KwResult{Graph $\mathcal{G}$ embedded on best value of $c$ for given $\Delta t$ for each bi-connected component of $\mathcal{G}$}
 find bi-connected components ${r}$ of $\mathcal{G}$\;
 Balance the forces between each $r$ such that for each $r$ $0 = \sum f_i$ \;
 Identify largest $r$ and Run auto-SETSe($\mathcal{G}_r$)\;
 $\textbf{G}_\textrm{vect} = \emptyset$\;
 \For{$r \in \mathcal{G}$}{
  Run auto-SETSe on all $r$\;
  $\textbf{G}_\textrm{vect} =\textbf{G}_\textrm{vect}+ \mathcal{G}_\textrm{r} $\;
 }

  $\mathcal{G}_\textrm{abs} = \textrm{abs\_elevation}(\textbf{G}_\textrm{vect})$\;

\Return $\mathcal{G}_\textrm{abs}$
 \caption{Bi-connected SETSe.}
  \label{algo:biconnSETSe}
\end{algorithm}

\begin{algorithm}
\SetAlgoLined
\KwResult{A graph with all embeddings set relative to a single reference value}
 
 $\mathcal{G}_\textrm{abs} =\emptyset$\;
 $\textrm{Art}_\textrm{vect} = \emptyset$\;
  
 $\mathcal{G}_\textrm{new} =\mathcal{G}_\textrm{1}$ \#Origin block\;

 $n = 1$ \# The origin block is indexed as 1\;
  $t = 1$ \# vector of indices of blocks to be added
 
 \While{$n \leq$ number of blocks}{
  $\mathcal{G}_\textrm{abs} =\mathcal{G}_\textrm{abs} + \mathcal{G}_\textrm{new}$\;
  
   $\textrm{Art}_\textrm{vect} := \textrm{Art}_\textrm{vect} + \left \{ n_\textrm{art} \in \mathcal{G}_\textrm{new} \right \}$\;
  $\textbf{G}_\textrm{vect}  = \textbf{G}_\textrm{vect} \setminus  t$ \#remove targeted blocks from block list \;
  
  $n = n+1$\; \#increment the index
  
  $t = \textrm{Index}(\left \{x \subset \textbf{G}_\textrm{vect} : \textrm{Art}_\textrm{vect} \left [ n \right ]   \in x \right \})$\;
  
  $\mathcal{G}_\textrm{new} = \emptyset$
  
  \For{$x \in t$}{
    
    subtract the elevation of node $\textrm{Art}_\textrm{vect} \left [ n \right ] $ in $\mathcal{G}_x$ from all nodes in $\mathcal{G}_x$ then add the elevation of $\textrm{Art}_\textrm{vect} \left [ n \right ] $ in $\mathcal{G}_\textrm{abs}$ to all nodes in $\mathcal{G}_x$.
    
    $\mathcal{G}_\textrm{new} = \mathcal{G}_\textrm{new} + \mathcal{G}_x$
    
 }  
  
 }
 
 \Return $\mathcal{G}_\textrm{abs}$

 \caption{abs\_elevation. Combine all bi-connected components into a single structure.}
  \label{algo:abselevation}
\end{algorithm}

\begin{figure}
    \centering
    \includegraphics{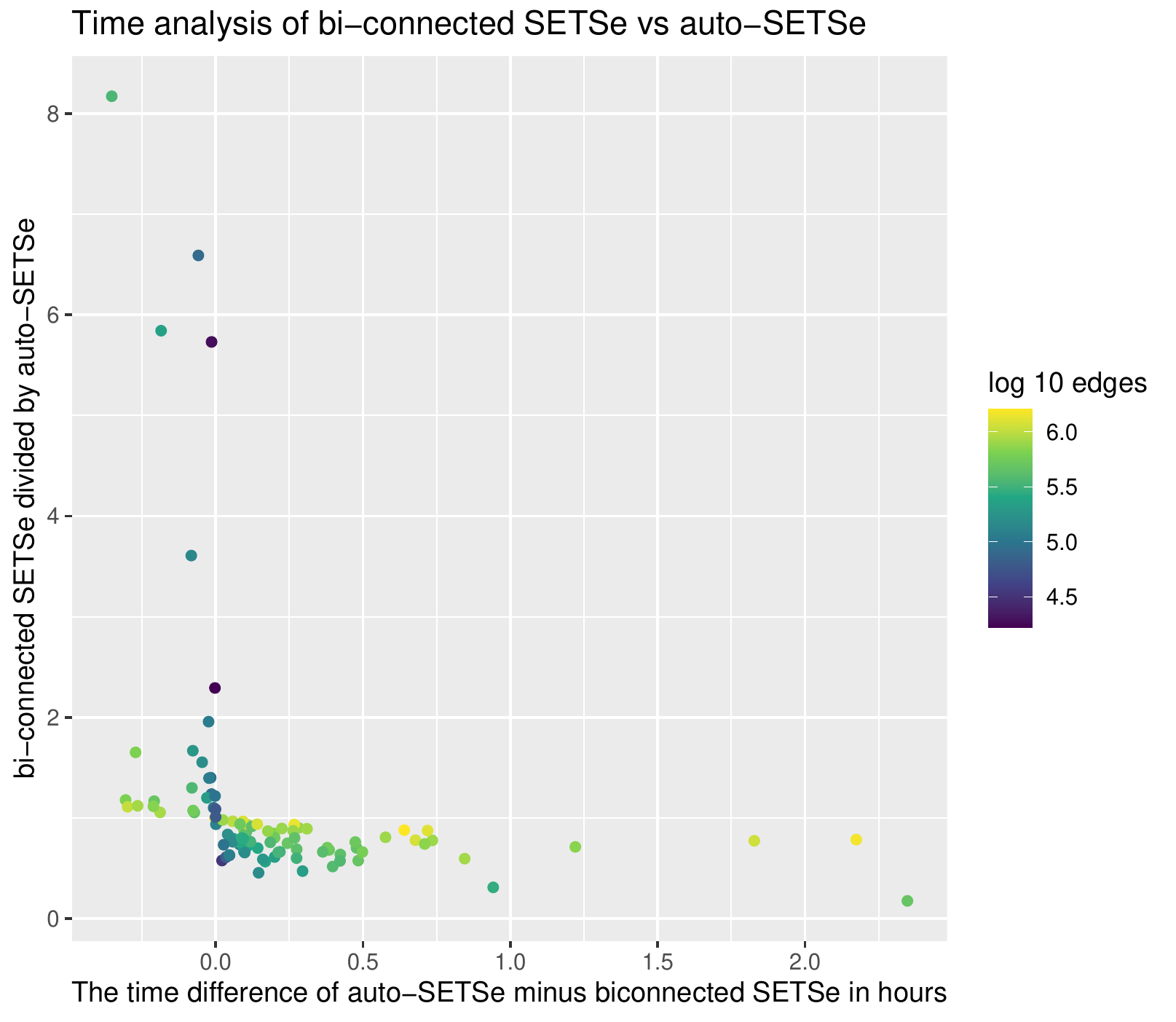}
    \caption{Although the bi-connected version of SETSe is not guaranteed to be faster than auto-SETSe, across multiple networks the bi-connected method is substantially faster, this is particularly true for larger networks. }
    \label{fig:bicomp_vs_auto}
\end{figure}

The network Caltech36 is in fact over 2 times slower using the bi-connected method (although still only 12 seconds). However, nodes with only 1 neighbour now close to no static force (2e$^{-10}$N), and the over all network has 20\% less static force when using the biconnected method when compared to auto-SETSe. 
Figure \ref{fig:bicomp_vs_auto} shows the difference in convergence time for the Facebook 100 dataset using auto-SETSE and the bi-connected method. On the $y$-axis, the figure shows the time taken using the bi-connected over the time taken for the auto-SETSe version. The $x$-axis shows the difference in hours. It can be seen that the bi-connected version is faster than auto-SETSe 73\% of the time. However, when the bi-connected algorithm is slower, the reduction in speed is small in absolute terms. The worst case increases the time by 21 minutes, and the best case decreases the time by 2.3 hours. Overall, using the bi-connected algorithm was a modest 13\% faster than using auto-SETSe alone, saving approximately 21.6 hours of computation time. If both algorithms are allowed to run across multiple cores, the bi-connected can be several times faster, from the user perspective. Both approaches successfully converged over 95\% of the networks, but needed a small change to the number of iterations used when finding the coefficient of drag for the remaining networks  (2000 to 1000 hyper iterations). A detailed explanation of when the bi-connected version is faster is beyond the scope of this paper; however, it appears loosely related to the size of the network and the number of bi-connected components.

\printbibliography